\documentclass[useAMS,usenatbib]{mn2e}
\usepackage{float,graphics,epsfig,array,amsmath,amssymb,times}

\newcommand{\be}{\begin{equation}}  \newcommand{\ee}{\end{equation}}
  \newcommand{\ba}{\begin{eqnarray}}
\newcommand{\bm}[1]{\mbox{\boldmath{$#1$}}}

\title[Bayesian galaxy shape measurement]
{Bayesian Galaxy Shape Measurement for Weak Lensing Surveys -III. 
Application to the Canada-France-Hawaii Telescope Lensing Survey.
}
\author[L.~Miller et~al.]
{L.~Miller$^1$, 
C.~Heymans$^2$, 
T.D.~Kitching$^2$, 
L.~Van Waerbeke$^3$,
T.~Erben$^4$,\newauthor
H.~Hildebrandt$^{3,4}$, 
H.~Hoekstra$^{5,6}$, 
Y.~Mellier$^7$, 
B.T.P.~Rowe$^{8,9,10}$,
J.~Coupon$^{11}$,\newauthor
J.P.~Dietrich$^{12}$,
L.~Fu$^{13}$,
J.~Harnois-D\'{e}raps$^{14,15}$, 
M.J.~Hudson$^{16,17}$,
M.~Kilbinger$^{18,19,20,7}$,\newauthor
K.~Kuijken$^{5}$,
T.~Schrabback$^{4,21,5}$,
E.~Semboloni$^{5}$,
S.~Vafaei$^{3}$
and
M.~Velander$^{1,5}.$
\\
$^1$Department of Physics, Oxford University, Keble Road, Oxford OX1 3RH, UK.\\ 
$^2$Scottish Universities Physics Alliance, Institute for Astronomy, University of Edinburgh, Royal Observatory, Blackford Hill, Edinburgh, EH9 3HJ, UK. \\ 
$^3$Department of Physics and Astronomy, University of British Columbia, 6224 Agricultural Road, Vancouver, V6T 1Z1, BC, Canada.\\  
$^4$Argelander Institute for Astronomy, University of Bonn, Auf dem H{\"u}gel 71, 53121 Bonn, Germany.\\
$^5$Leiden Observatory, Leiden University, Niels Bohrweg 2, 2333 CA Leiden, The Netherlands.\\
$^6$Department of Physics and Astronomy, University of Victoria, Victoria, BC V8P 5C2, Canada.\\
$^7$Institut d'Astrophysique de Paris, Universit\'e Pierre et Marie Curie - Paris 6, 98 bis Boulevard Arago, F-75014 Paris, France.\\
$^{8}$Department of Physics and Astronomy, University College London, Gower Street, London WC1E 6BT, U.K.\\
$^{9}$Jet Propulsion Laboratory, California Institute of Technology, 4800 Oak Grove Drive, Pasadena CA 91109, USA.\\
$^{10}$California Institute of Technology, 1200 E California Boulevard, Pasadena CA 91125, USA.\\
$^{11}$Institute of Astronomy and Astrophysics, Academia Sinica, P.O. Box 23-141, Taipei 10617, Taiwan.\\
$^{12}$Department of Physics, University of Michigan, 450 Church St., Ann Arbor, MI 48109, USA.\\
$^{13}$Key Lab for Astrophysics, Shanghai Normal University, 100 Guilin Road, 200234, Shanghai, China. \\
$^{14}$Canadian Institute for Theoretical Astrophysics, University of Toronto, M5S 3H8, Ontario, Canada.\\
$^{15}$Department of Physics, University of Toronto, M5S 1A7, Ontario, Canada.\\
$^{16}$Department of Physics and Astronomy, University of Waterloo, Waterloo, ON, N2L 3G1, Canada.\\
$^{17}$Perimeter Institute for Theoretical Physics, 31 Caroline Street N, Waterloo, ON, N2L 1Y5, Canada.\\
$^{18}$CEA Saclay, Service d'Astrophysique (SAp), Orme des Merisiers, B\^at 709, F-91191 Gif-sur-Yvette, France.\\
$^{19}$Excellence Cluster Universe, Boltzmannstr. 2, D-85748 Garching, Germany.\\
$^{20}$Universit\"ats-Sternwarte, Ludwig-Maximillians-Universit\"at M\"unchen, Scheinerstr.~1, 81679 M\"unchen, Germany.\\
$^{21}$Kavli Institute for Particle Astrophysics and Cosmology, Stanford University, 382 Via Pueblo Mall, Stanford, CA 94305-4060, USA.\\
}

\begin{document}

\pagerange{\pageref{firstpage}--\pageref{lastpage}} \pubyear{2011}

\maketitle

\label{firstpage}

\begin{abstract}
A likelihood-based method for measuring weak gravitational lensing shear in
deep galaxy surveys is described and applied to the
Canada-France-Hawaii Telescope (CFHT) Lensing Survey
(CFHTLenS). CFHTLenS comprises 154\,deg$^{2}$ of multicolour optical
data from the CFHT Legacy Survey, with lensing measurements being made
in the $i'$ band to a depth $i'_{\rm AB}<24.7$, for galaxies with
signal-to-noise ratio $\nu_{\rm SN} \ga 10$.  The method is based on
the {\em lens}fit algorithm described in earlier papers, but here we
describe a full analysis pipeline that takes into account the
properties of real surveys.  The method creates pixel-based models of
the varying point spread function (PSF) in individual image exposures.
It fits PSF-convolved two-component (disk plus bulge) models, to measure the
ellipticity of each galaxy,
with bayesian marginalisation over model nuisance parameters
of galaxy position, size, brightness and bulge fraction.  The method
allows optimal joint measurement of multiple, dithered image
exposures, taking into account imaging distortion and the alignment of
the multiple measurements. We discuss the effects of noise bias on the
likelihood distribution of galaxy ellipticity.  Two sets of image
simulations that mirror the observed properties of CFHTLenS have been
created, to establish the method's accuracy and to derive an empirical
correction for the effects of noise bias.
\end{abstract}

\section{Introduction}
Weak gravitational lensing allows measurement of the matter content of the Universe,
which, when used in combination with other probes, should enable tests of dark energy models
of the Universe and tests of general relativity on cosmological scales.  However, despite
the detection of weak lensing around galaxy clusters more than 20 years ago \citep[e.g][]{tyson90a}
and the detection of cosmological large-scale lensing (`cosmic shear') more than 10 years ago
\citep{bacon00a, vanwaerbeke00a, wittman00a}, it has proved
difficult to make measurements that are free from systematic error
at the level required to measure cosmological parameters accurately.
To date, imaging data obtained from the Hubble Space Telescope (HST) has been the most successful in
this respect, but has yielded accurate cosmological results only
relatively recently \citep[e.g.][]{schrabback10a}.  Extension of these studies to large
sky areas is currently only feasible from the ground, but then the weak lensing measurements
need to be made using images of faint galaxies that are only marginally resolved 
and for which the lensing signal needs to be disentangled from the effects of an 
image PSF which varies both with position within an image and between images.

\begin{figure}
\resizebox{0.48\textwidth}{!}{
\rotatebox{-90}{
\includegraphics{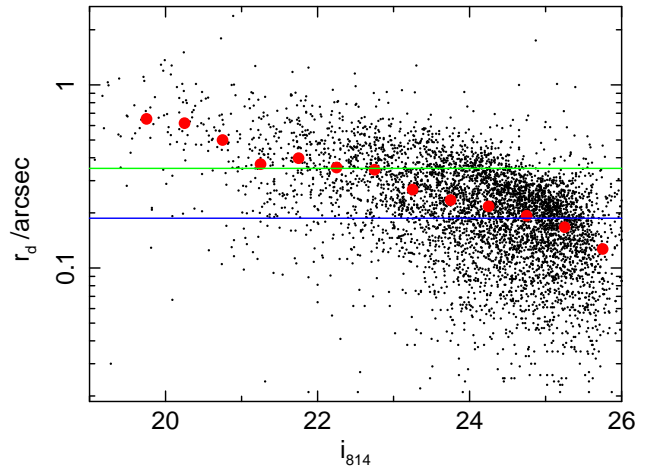}
}}
\caption{
Fitted disk semi-major axis scalelength, plotted as a function of $i'$-band magnitude, for galaxies
identified as being disk-dominated, from the fits of
\citet{simard02a} to HST WFPC2 data. The large filled circles indicate
the median size measured in bins of apparent magnitude.  The horizontal lines indicate the
CFHT MegaCam pixel size of $0.187''$ (lower line) 
and a typical CFHTLenS PSF half-width half maximum of $0.35''$ (upper line).
\label{fig:simard}
}
\end{figure}

Fig.\,\ref{fig:simard} shows the sizes of disk galaxies in the `Groth Strip', 
as measured by \citet{simard02a} using data from the 
HST Wide Field Planetary Camera\,2 (WFPC2).  These data are used in Appendix\,\ref{appendixB} to obtain
an estimate of the bayesian prior distribution of galaxy size used in this paper,
however it is useful to show the distribution here to illustrate the level of difficulty 
imposed by ground-based measurements. The figure shows the fitted values of semi-major
axis disk exponential scalelength obtained for galaxies which had fitted bulge-to-total
flux ratios (B/T) less than 0.5\footnote{
A population of objects that were fitted by \citet{simard02a} with
scalelengths less than $0.02''$ have been excluded: these very small scalelengths are not correlated with
apparent magnitude and are likely fits to stars or are otherwise spurious measurements, as noted by
\citeauthor{simard02a}.}. 
We have measured the median values of that scalelength
in bins of $i$-band (WFPC2 filter F814) apparent magnitude, excluding values $r_d<0.02''$,
and these are shown as large filled circles in Fig.\,\ref{fig:simard}.  
In this paper, we describe how we have made weak lensing measurements using 
ground-based data from
the Canada-France-Hawaii Telescope (CFHT) MegaCam camera \citep{boulade03a}
obtained for the CFHT Legacy Survey (CFHTLS) and analysed as the 
CFHT Lensing Survey (hereafter CFHTLenS): Fig.\,\ref{fig:simard} also shows the 
MegaCam typical pixel size of $0.187''$ and a typical CFHTLS PSF half-width half-maximum 
(HWHM) of
$0.35''$.  It can be seen that at the faintest magnitudes (CFHTLenS analysed galaxies to
$i'_{\rm AB}<24.7$) the median semi-major axis exponential 
scalelength of galaxies is comparable to the pixel size, and that very few galaxies have 
scalelengths that are larger than the PSF HWHM.  Given that the faint galaxies are also
noisy, with the faintest galaxies having signal-to-noise ratio as low as 7, it is clear
that highly accurate methods need to be adopted for measuring the systematic distortions
of galaxies at the level of a few percent in ellipticity that are expected from weak
lensing.

We describe in this paper a bayesian model-fitting method aimed at
measuring the shapes of faint galaxies in optical galaxy surveys.
Current weak lensing surveys such as CFHTLenS comprise about $10^7$
galaxies, so one design aim of the algorithms is that they should be
fast.  However, as survey sizes increase, the requirements on
acceptable levels of non-cosmological systematics signals become more
stringent, meaning that the new generation of measurement algorithms
must be simultaneously faster and more accurate than previous methods.
The principles that were adopted for measuring galaxy ellipticity are
described by \citet[][hereafter Paper\,I]{miller07a}, and implemented in
the `{\em lens}fit' algorithm.  Tests on simulated images 
of a previous implementation of the basic shear-measurement algorithm are described by
\citet[][Paper\,II]{kitching08a}.

This paper describes a number of significant improvements to the algorithm which proved necessary
for working on data from a real survey, rather than simplified simulations.
\begin{enumerate}
\item PSF modelling took account of spatial variation of the PSF and possible discontinuities
between CCDs in the mosaic camera (MegaCam, \citealt{boulade03a}, in the case of CFHTLenS).
\item Galaxy measurement used multiple, dithered exposures, optimally combined in the measurement process, taking account of pointing and camera distortion variations and varying image quality.
\item Camera distortion was measured directly from images and corrected in the measurement process.
\item Fitted galaxy models allowed joint fitting of bulge and disk components.
\item Full bayesian marginalisation over nuisance parameters was carried out.
\item Additional measures were used to account for bad and saturated pixels and columns, cosmic rays, 
image ghosts, blended stars and galaxies and recognition of morphologically complex galaxies. 
\end{enumerate}
Information on these various improvements is described below. Section\,2 describes CFHTLenS, Sections\,3--6 provide details of the method and the shear measurement pipeline, Section\,7 provides an overview of
tests for systematic errors, which are described in more detail by \citet{heymans12a}, 
and Section\,8 describes
extensive image simulations that have been used to test the method and to calibrate the effects
of noise bias.  Appendices\,A and\,B provide additional details of the models and priors used in the method, 
and a mathematical illustration of the effect of noise bias is given in Appendix\,\ref{appendixC}.

\section{The Canada-France Hawaii Telescope Lensing Survey}\label{sec:cfhtlens}
The data used in this paper are from the CFHT Wide Legacy Survey\footnote{www.cfht.hawaii.edu/Science/CFHTLS},
analysed by the CFHTLenS team\footnote{www.cfhtlens.org}. The CFHTLS Wide consisted of 171 pointings
covering a sky area of 154\,deg$^2$ in five bands, $u^*g'r'i'z'$, 
with the MegaCam camera \citep{boulade03a}, which has a field of view of approximately 1\,deg$^2$.
The pointings were distributed
over four separate patches on the sky, and
each patch was contiguous, covering an area of 63.8, 22.6, 44.2 and 23.3\,deg$^2$, respectively for the
W1, W2, W3 and W4 patches.  Each pointing comprised typically seven $i'$-band exposures, often with many 
of those taken during the same night.
In some fields, one or two exposures were judged to be of too poor quality, based on the seeing or apparent
degree of PSF variation, reducing the number of
usable exposures to 5 or 6.  In some other fields, two entire sets of exposures were obtained
if there was concern about the quality of the first set, but in a number of cases all 14 exposures
were subsequently judged to be usable and were included in the CFHTLenS analysis.
Exposures were dithered by up to $90''$ in declination and $24''$ in right ascension in patches W1, W3 and W4,
 and by up to $400''$ in W2.

The CFHTLenS analysis of the data was designed and optimized for weak
lensing science, both photometry and shape measurements. The photometric analysis of the
survey is described by \citet{hildebrandt12a} and \citet{erben12a}.
Galaxy shapes have been measured on the $i'$ band which reaches a $5\sigma$
point source limiting magnitude of $i'_{AB} \sim 25.5$. A magnitude limit of $i'_{AB} < 24.7$
was imposed for shape measurement, noting that photometric redshift accuracy also is poor at
fainter magnitudes than this limit \citep{hildebrandt12a}.
Exposures obtained prior to June~2007 were obtained with the $i'$-band filter MP9701.  After October~2007
an $i'$-band filter with a slighter different spectral response, MP9702, was used.  

All fields had a full-width half-maximum (FWHM) seeing condition which was required to be better
than $0.8''$. Unlike previous lensing analyses, galaxy shape measurement was
performed on the individual exposures, photometrically calibrated with Elixir \citep{magnier04a}.
The CFHTLenS data were then processed with the {\sc theli} pipeline \citep[see][]{erben05a}.
The specific algorithms and procedures were described by \citet{erben09a}.
In that work, the astrometric and
photometric calibration was made independently in each 1~deg$^2$ MegaPrime pointing.
The most important refinement to the calibration for CFHTLenS was a simultaneous
treatment of all data from a CFHTLenS patch. 
This typically led to an improvement of a factor 2--3
in astrometric and photometric accuracy compared with the earlier analysis.
Further details of the construction of the survey are given by \citet{heymans12a}
and \citet{erben12a}.

\section{The shear measurement method}

\subsection{Principles}
Key points to emphasise about the approach of bayesian model-fitting are
\begin{enumerate}
\item
It enables a rigorous statistical framework to be established for the measurement,  
resulting in optimally-weighted use of the available data, regardless of variations in 
signal-to-noise ratio or PSF.
\item
Quantities other than ellipticity that form part of the problem may be marginalised over and, provided
an accurate prior is available, those uninteresting quantities may be eliminated from the problem without
bias.  The prime example is that of the unknown galaxy position: all other methods published to date
require a fixed centroid position for a galaxy, and centroid errors introduce spurious ellipticity, with
the possibility of bias \citep[e.g.][]{bernstein02a}.  In model-fitting we can treat the unknown galaxy position as a free parameter
and marginalise.
\item
The weak lensing signal is carried by the faintest galaxies, often the integrated signal-to-noise ratio may be
as low as 10.  A model-fitting approach adds extra information to the problem, 
such as prior knowledge of possible surface brightness distributions.
In principle this
leads to more precise measurements than a method that does not use such information, thereby allowing
measurements to lower signal-to-noise ratio.
\item
The models must be unbiased, however, in the sense that they should span the range of surface brightness
profiles of real galaxies, and the priors adopted for the parameters of those models should be an accurate reflection of the true distribution.
If these conditions are not satisfied, the model-fitting may be systematically biased \citep{voigt10a, bernstein10a}.
\item
A full bayesian approach should be able to correct for the effects of noise bias, and hence avoid
the need for any independent calibration.  We shall see in Section\,\ref{shear} that we
have not yet achieved that goal.
\end{enumerate}

Point (iv) is worth considering in more detail.  \citet{bernstein10a} has argued 
that model basis functions need to be truncated to some finite set to avoid the
fitting results being dominated by noise, and that truncation leads to bias \citep[see also][]{ngan09a}.
Thus, even when specific physical models are not assumed, the use of
any so-called `model-independent' basis set also results in biased results,  unless the
truncated basis set employed is capable of accurately modelling the full observed range of surface
brightness profiles.  In methods that use simple statistical measures of the data, such as
moments (e.g \citealt{kaiser95a}, hereafter KSB, or \citealt{melchior11a}) 
it is necessary to use weighted
moments because of image noise, and the use of a finite set of weighted moments can lead
to a bias similar to the use of truncated basis sets.  
Thus, all methods that aim to measure galaxy shapes from noisy, PSF-convolved data have
the risk of producing biased results, and choosing a so-called `model-independent' method does
not alleviate this concern \citep[see also][]{melchior12a}.
The approach adopted here is that if the model basis sets are a close
match to the true galaxy surface brightness distributions, and in particular if the PSF-convolved
model distributions span the range of true PSF-convolved distributions, then it should be possible
to avoid model truncation bias.  By choosing model surface brightness distributions that match
the bulge and disk components of galaxies at low redshift, we are making a choice of a basis set
that should minimise bias.  That does leave open the question of whether such models can span the
range of galaxy surface brightness distributions at high redshift, given the expected 
morphological evolution of the galaxy population, but whether or not the fitting of composite
disk and bulge models to galaxies at high redshift results in bias has not yet been established.
This concern should be explored in future work, especially given the ever-increasing accuracy required
for future surveys.

The models fitted in CFHTLenS comprised two-component bulge plus disk galaxies (Appendix\,\ref{appendixA}).  
There were seven free parameters:
galaxy ellipticity, $e_1$, $e_2$; galaxy position $x$, $y$; galaxy size $r$; galaxy flux; and bulge fraction.
Ultimately we are interested only in the ellipticity, so the other parameters were regarded as nuisance
parameters and marginalised over.  Bayesian marginalisation was used, adopting prior distributions
for each parameter, which are described in Appendix\,\ref{appendixB}.
With the given priors, each parameter was marginalised over either analytically, if possible, or 
otherwise numerically.
The bulge fraction marginalisation was analytic, 
galaxy flux and size were numerically marginalised over by sampling multiple
values, fitting a function and integrating.  Galaxy position was numerically marginalised over 
using the Fourier method described in Paper\,I: the process of model-fitting
is equivalent to cross-correlation of the model with the data, and marginalisation over unknown position
is equivalent to integrating under the peak of the cross-correlation function.  
Thus the likelihood surface was obtained as a function of only the apparent 
galaxy two-component ellipticity.  Finally,
that likelihood surface was converted to weighted estimates of ellipticity 
for use in shear measurement (Section\,\ref{shear}).  
The following sections give more detail of the galaxy models and of the
calculation of the
likelihood function and its marginalisation.

\subsection{Galaxy models}\label{galaxymodels}
Adopting the same conventions as in Paper\,I, observed galaxy ellipticity $\bmath{e}$ is
related to the intrinsic galaxy ellipticity $\bmath{e^s}$ by
\begin{equation}
\bmath{e} = \frac{\bmath{e^s} + \bmath{g}}{1 + \bmath{g^{*}e^s}}
\label{eqn:shear}
\end{equation}
\citep{schramm95a,seitz97a}, where $\bmath{e}$ and $\bmath{g}$ are represented as complex variables,
$\bmath{g}$, $\bmath{g^{*}}$ are the reduced shear and its complex conjugate
respectively, and $\bmath{e}$ is defined in terms of the major and minor axes and orientation
$a, b, \theta$ respectively, as $\bmath{e} = (a-b)/(a+b)\exp(2i\theta)$.
In this formalism, we expect
\begin{equation}
\left\langle \bmath{e} \right\rangle = \bmath{g}
\label{eqn:weakshear}
\end{equation}
for an unbiased sample with $g<1$, where $\left\langle \bmath{e^s} \right\rangle = 0$. 
Note that this result differs from the other commonly used
formalism which instead uses ellipticity, or polarisation, defined as 
$\bmath{e} = (a^2-b^2)/(a^2+b^2)\exp(2i\theta)$,
and which requires measurement of the distribution of galaxy ellipticities in order to
calibrate the response to lensing shear \citep[e.g.][]{bernstein02a}.

The model-fitting approach to shear measurement of Paper\,I was adopted, but extended
to include multiple galaxy components, which we assumed to comprise a disk and bulge
component.  In reality, galaxies may have more complex morphologies than are
describable by this simple phenomenological model.  In non-bayesian methods,
models of greater complexity than required by the data lead to `overfitting' and its
associated biases \citep[e.g][]{bernstein10a}.
In principle, a bayesian method
could include additional components and model parameters, provided that prior probability
distributions are available, so that parameters not of interest to the shear measurement
may be marginalised over.  In practice, we wish to avoid using models of greater complexity
than are required by the data, particularly as more complex models increase the 
computational time required to make the measurements.  In CFHTLenS, early experiments
with the systematics tests described in Section\,\ref{systematics} and by \citet{heymans12a} indicated
that significant improvements could be obtained by extending the single-component model of
Paper\,I to a simple two-component galaxy model in which only one additional free parameter
was allowed.

The generation of the galaxy models is described in Appendix\,\ref{appendixA}.
We assumed that a galaxy survey may contain
two fundamental galaxy types. A fraction $f_B$ were bulge-dominated galaxies 
with a single morphological component, which was
assumed to be described by a de Vaucouleurs profile (S\'{e}rsic index 4).
A fraction $(1-f_B)$ were disk galaxies which 
comprised both a bulge component and a disk component.  The disk component was assumed to be a pure
exponential, S\'{e}rsic index 1, with the bulge component again having S\'{e}rsic index 4.  
The model components were truncated in surface brightness at a major axis 
radius of 4.5 exponential scalelengths (disk component) or 4.5 half-light radii (bulge component),
which was both 
computationally convenient and also in general agreement with observations of disk galaxies
\citep[e.g.][]{vanderkruit82a}.  

To avoid prohibitive computational cost by introducing another model parameter, which would
not be analytically marginalisable, the half-light radius of the bulge was set equal to the exponential
scalelength of the disk for the composite disk-dominated galaxies (i.e. the bulge half-light radius 
was 0.6 times the value of the disk half-light radius). In reality, galaxies display a wide
range in their ratios of the sizes of these components, although that distribution itself depends on the
assumed surface brightness profiles adopted \citep{graham08a}.  The adopted ratio of bulge half-light radius 
to exponential scalelength is towards the upper end of the range encountered in the literature.  
As the CFHTLenS galaxies are
generally only marginally resolved, smaller bulges would have half-light radii significantly smaller than
the pixel scale, and the final PSF-convolved models were largely insensitive to the choice
of this ratio, although we did find that significantly smaller values of bulge half-light radius
led to larger PSF-galaxy cross-correlations (see Section\,\ref{systematics}) for the smallest fitted galaxies,
and were therefore disfavoured.

Given the low signal-to-noise ratio and the small number of pixels
associated with the faintest galaxies in the CFHTLenS data, a more complex set of galaxy models, 
with more free parameters, was not justified.
We might ask, however, whether the particular models chosen are the `best', in the sense of being
the least biased, or whether some other choice might be better.  
To answer such a question, we need to have a good understanding of the actual population of galaxies
being measured.  Recently,
attempts have been made to create realistic simulations of the galaxy population based on
high-resolution HST observations \citep{mandelbaum12a}, and these form the basis of the 
GREAT3 challenge simulations\footnote{great3challenge.info}.  In the current paper,
we take the view that the bulge-plus-disk parameterisation is known to fit well
to the local galaxy population, and therefore the surface brightness distributions
of such models are already a necessary ingredient of any successful models.  
Even though high-redshift
galaxies tend to have more complex morphologies than local galaxies, the effects of
convolution with the PSF may render such differences unimportant in ground-based data of
CFHTLenS quality, where the high-redshift galaxies are only marginally resolved.
We discuss in Section\,\ref{modelbias} the possible amplitude of biases that might arise
from inappropriate model choice.
The requirements for larger, future surveys, with correspondingly
tighter requirements on the acceptable level of systematic biases, will need to be
established through more extensive simulations.

\subsection{The likelihood function and marginalisation over nuisance parameters}\label{likel}
We define the likelihood $\mathcal{L}$ and the $\chi^2$ goodness-of-fit statistic as
\begin{equation}
\chi^2 \equiv -2\log\mathcal{L} = \sum_i \left[\frac{y_i-SBg_i-S(1-B)f_i}{\sigma_i}\right]^2,
\label{eqn:chisq}
\end{equation}
where $y_i$ is the image data value of pixel $i$, $\sigma_i$ is the statistical
uncertainty of that data value, $f_i$ is the disk component model value for that pixel, 
convolved with the PSF,
$g_i$ is the bulge component model amplitude, 
convolved with the PSF,
$B$ is the bulge fraction defined as the
ratio of the flux in the bulge component to the total flux (often written B/T) and
$S$ is the total galaxy flux.  
The pixel noise is assumed to be 
stationary and uncorrelated, which is appropriate for 
shot noise in CCD detectors in the sky-noise limit.  
Bright galaxies may make a significant contribution to photon shot noise, but the 
resulting non-stationary noise would then invalidate the {\em lens}fit Fourier-space approach
which is described in Paper\,I and used below.
Hence this algorithm is appropriate for model-fitting to faint galaxies
in the sky-limited regime. 
The method can easily be generalised to the case where the noise is
stationary but correlated between pixels, but we do not consider that generalisation here.

The prior distribution of bulge fraction $B$ for the disk galaxy population
was assumed to be a truncated gaussian distribution,
\begin{equation}
\mathcal{P}(B) = \mathcal{P}_0 \exp\left[-\left(B-B_0\right)^2/\left(2{\sigma_{\rm B}}^2\right)\right] \hspace*{3mm} {\rm 0\le B < 1}
\label{eqn:bulgeprior}
\end{equation}
where $\mathcal{P}_0$ was a normalising factor and $B_0$ and $\sigma_{\rm B}$ were parameters of the prior distribution. 
The gaussian form of the prior was a convenient choice to allow easy marginalisation
over $B$, given the quadratic dependence of $\mathcal{L}$ on $B$.
We adopted $B_0 = 0$, $\sigma_{\rm B} = 0.1, f_B=0.1$ as a reasonable representation
of the distribution of bulge fraction found at low and high redshift (e.g.
\citealt{schade96a}, \citealt{simard02a}). 
We marginalised over the gaussian part of the prior by
collecting terms in equation\,\ref{eqn:chisq} and multiplying by the prior, writing the 
log of the posterior probability density for the disk population, $\log P_{\rm disk}$, as
\begin{eqnarray}
\nonumber
-2 \log P_{\rm disk} &= & \textstyle\sum y^2/\sigma^2 + 2B^2\theta^2 + 2Br_0 \\
&+& \textstyle S^2\sum f^2/\sigma^2 - 2S \sum fy/\sigma^2 +B_0^2/\sigma_{\rm B}^2
\label{eqn:chisqexpand}
\end{eqnarray}
where
\begin{eqnarray}
\nonumber
2\theta^2 &=& 
\textstyle
S^2\left[\sum f^2/\sigma^2 + \sum g^2/\sigma^2 -2\sum fg/\sigma^2 \right]+ 1/\sigma_{\rm B}^2,
\\
\nonumber
r_0 &=& \textstyle S\left(\sum fy/\sigma^2 - \sum gy/\sigma^2\right)\\
\nonumber
&+& \textstyle S^2\left(\sum fg/\sigma^2 - \sum f^2/\sigma^2\right)
- B_0/\sigma_{\rm B}^2,
\end{eqnarray}
and where the summations are understood to be over the pixels $i$.  
We marginalised
over $B$ to obtain the log of the marginalised disk posterior probability density as
\begin{eqnarray}
\nonumber
\lefteqn{
-2\log P_{\rm disk} =
\textstyle S^2\sum f^2/\sigma^2 - 2S \sum fy/\sigma^2
-\frac{r_0^2}{2\theta^2}}\\
&+&2\log\theta
-2\log\left[{\rm erf}\left(\theta+\frac{r_0}{2\theta}\right)-{\rm erf}\left(\frac{r_0}{2\theta}\right)\right],
\label{eqn:posterior}
\end{eqnarray}
neglecting any normalisation terms that do not depend on a galaxy's model parameters.

The bulge-fraction-marginalised total probability $P = (1-f_{\rm B})P_{\rm disk} + f_{\rm B}P_{\rm  bulge}$
then needed to be marginalised over both flux and position.  
To marginalise over galaxy position at a given model galaxy flux, 
we followed an improved version of the method described in Paper\,I.
For stationary noise, the terms $\sum f^2/\sigma^2$ and $\sum g^2/\sigma^2$ are invariant under shifts of the model
position, and following Paper\,I, the terms $\sum fy/\sigma^2$ and $\sum gy/\sigma^2$ may be regarded as the
spatial cross-correlation of each model component with the data, so we may write 
\begin{equation}
\textstyle
h_f(\bmath{x}) = \sum fy/\sigma^2, \hspace*{1cm}
h_g(\bmath{x}) = \sum gy/\sigma^2,
\end{equation}
where the cross-correlations $h_f$ or $h_g$ depend on the vector position shift $\bmath{x}$ of the model.
As pointed out in Paper\,I,
to marginalise over $\bmath{x}$ we need to adopt a prior $\mathcal{P}(\bmath{x})$: 
{\em a priori}, the position of the galaxy is unknown, and is equally likely to exist
at any position on the sky, so we adopt a uniform prior.  Any other choice would lead to
a bias in the position of the galaxy, and hence a bias in the measured ellipticity, arising
from the correlation between fitted ellipticity and position.  
However, as discussed in Paper\,I, $\mathcal{L}\rightarrow$constant as
$|\bmath{x}|\rightarrow\infty$ and the marginalised likelihood would not be finite.
This problem arises because, no matter how large a pixel value, it always has a finite
chance of being due to random noise, with the true galaxy being positioned elsewhere.
At large position offsets, $\bmath{x} \rightarrow \infty$, $h_f, h_g \rightarrow 0$
and the posterior probability density tends towards a `pedestal' value 
$P_0 \equiv (1-f_B)P_{0,{\rm disk}} + f_B P_{0, {\rm bulge}}$, where
\begin{eqnarray}
\nonumber
-2\log P_{0,{\rm disk}} &=& 
\textstyle S^2\sum f^2/\sigma^2 -\frac{r_0'^2}{2\theta^2} +2\log\theta\\
\nonumber
&-&2\log\left[{\rm erf}\left(\theta+\frac{r_0'}{2\theta}\right)-{\rm erf}\left(\frac{r_0'}{2\theta}\right)\right],\\
\nonumber
r_0' &=& \textstyle S^2\left(\sum fg/\sigma^2 - \sum f^2/\sigma^2\right)
- B_0/\sigma_{\rm B}^2,\\
\nonumber
-2\log P_{0,{\rm bulge}} &=& \textstyle S^2 \sum g^2/\sigma^2.
\end{eqnarray}
Thus we need to identify a maximum bound for the position offset over which to marginalise.

As argued in Paper\,I, 
galaxies generally have smooth centrally-concentrated surface brightness 
distributions which are convolved with centrally-concentrated PSFs in
an observed image (the assumption of a smooth convolved galaxy profile is a good approximation for
galaxies that are not well resolved). 
The model is also smooth,
centrally concentrated and convolved with the same PSF.  
As in the derivation of the central limit theorem, such a cross-correlation
should be well represented by a two-dimensional gaussian distribution,
and hence we approximated $\log P$ as being a 2D gaussian added to the
pedestal value $\log P_0$.  To marginalise over the galaxy's position, we adopted 
a more exact integration of this form than assumed in Paper\,I,
and integrated out to some maximum position uncertainty, $r_{\rm max}$.  To establish the
value of $r_{\rm max}$, we considered that the only information that we have about the
galaxy is on the images being analysed: the galaxy probably was initially detected,
and its position measured, with some uncertainty, from the same data being used to measure the shape.
We are seeking to use the data in this region of the image to measure the shape of a galaxy that 
exists at this location, not of any arbitrary galaxy anywhere else in the image: although the
likelihood is non-zero at large position offsets, such a galaxy would not correspond to the
one which had been detected.  Accordingly we set $r_{\rm max}$ to be the position beyond which the
detection of this galaxy becomes statistically insignificant, which we define to be given by
$\Delta\log P = P(r_{\rm max})-P_0 < \Delta\chi^2_{\rm crit}/2$ and where we chose 
$\Delta\chi^2_{\rm crit} = 6$, corresponding to the 95\,percent confidence region for the location
of the galaxy.  The integral is given by an exponential integral
which was evaluated numerically using the GNU Scientific Library \citep{galassi09a},
rather than using the approximation of Paper\,I.

In Paper\,I, the flux marginalisation was carried out analytically
by noting the gaussian dependence of $P$ on $S$.  For the CFHTLenS analysis, it was decided
to adopt a prior distribution for $S$ which was given by a power-law in $S$: 
$\mathcal{P}(S) \propto S^{-\alpha}$ where $\alpha=2$ was chosen to match the flatter-than-Euclidean
slope of the faint $i$-band galaxy number counts \citep[e.g.][]{gabasch08a}.  The resulting posterior
probability distribution cannot then be analytically marginalised over $S$, so this was achieved
numerically, by evaluating $P$ at a set of points close to the expected maximum likelihood,
fitting a cubic spline function to the variation of $\log P$ with $S$ and numerically integrating
the spline-fitted values of $P$.

After these marginalisation steps, the marginalised likelihood had been reduced to being a
function of ellipticity and galaxy scalelength alone.  The final step was to adopt a prior for
the scalelength distribution (described in Appendix\,\ref{appendixB})
and to numerically marginalise over this parameter.  As for the
flux marginalisation, $\log P$ was measured for a set of values of scalelength $r$, a cubic
spline fitted to those values, and $P$ was numerically integrated with respect to $r$.  The
set of $r$ values measured included a point source, $r=0$, and it was ensured that the set of
$\log P$ values to be fitted varied smoothly towards $r=0$.  To distinguish between stars and galaxies,
objects were classified using an F-ratio test of the ratio of the 
chi-squared values of the best-fit galaxy ($r>0$) and the best-fit star ($r=0$) models, taking into account the
differing numbers of degrees of freedom of these two fits. A star classification was adopted if the
probability of the F-ratio exceeded a magnitude-dependent limit that was matched to visual classification of
objects\footnote{A bayesian procedure that took account of the relative probabilities of stars and 
galaxies as a function of magnitude could avoid needing to specify the value of the F-ratio statistic,
however given the very small numbers of stars expected in the faint galaxy, weak lensing, regime, $i \ga 23$,
this was not deemed to be an important improvement in this survey.}.
Of the objects that were fitted, typically 95\% were classified as galaxies, as expected at faint magnitudes.

\subsection{Sampling the likelihood surface}
In order to obtain unbiased estimates of shear, and to measure the statistical uncertainty in
ellipticity for each galaxy, we need to not only find the model ellipticity whose marginalised
probability is a maximum, but also we need to sample around the maximum to measure the
likelihood surface, at least in regions where the probability is not insignificant.  The ellipticity
surface is two-dimensional, and bounded by the requirement $|e|<1$, so it is relatively straightforward to
sample the surface appropriately.

However, galaxies of low signal-to-noise ratio or small size are expected to have broad likelihood surfaces,
whereas high signal-to-noise galaxies should have sharp, well-defined surfaces.  Thus the sampling
needed depends on the properties of each galaxy.  The algorithm adopted here was an adaptive grid
sampling, in which ellipticity measurements were first made on a coarse grid of ellipticity, at 
intervals of 0.16 in ellipticity.  At each sampled value, a full marginalisation over the other parameters
was made.  Then, the algorithm counted the number of sampled points above a threshold of 5 percent
of the maximum posterior probability, 
and if that number was below a minimum value of 30 points, the sampling 
in ellipticity was reduced by a factor 2, and the region around the maximum, only, sampled at
the higher resolution, until either 30 points had been measured above the threshold,
or until a resolution of 0.02 in ellipticity was reached (for the faint galaxies that dominate the weak
lensing signal this limiting resolution in ellipticity was adequate, and is much smaller than the 
`shape noise' caused by the broad distribution of intrinsic ellipticity).

In a later improvement, two passes were made through the ellipticity surface, the first to find 
only the maximum using the adaptive sampling algorithm, the second to map out the surrounding
region.  This algorithm was found to be more efficient (in that the fraction of sampled points 
that were retained above the threshold was higher) but was not implemented in time for the
CFHTLenS analysis reported here.  Monte-Carlo Markov Chain (MCMC) methods could also be used, but
given the low dimensionality and finite bounds of the parameter space, are unlikely to be
significantly more efficient than this adaptive sampling algorithm. 

\begin{figure*}
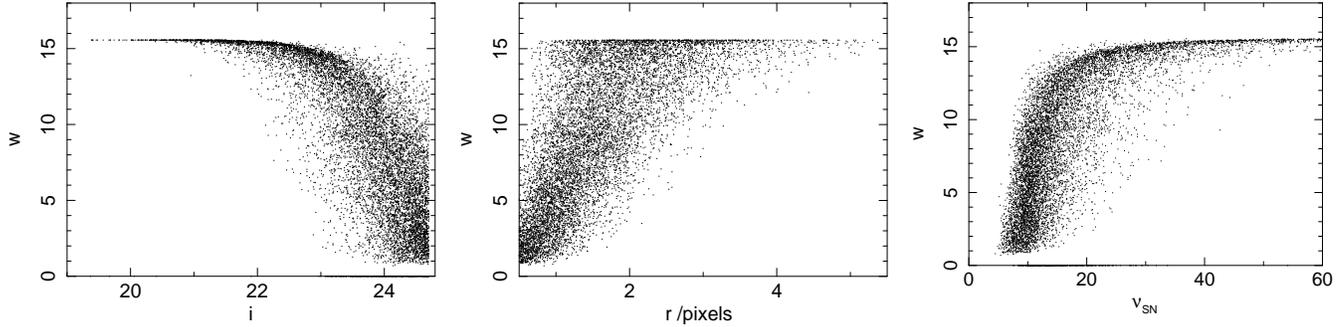

\resizebox{0.32\textwidth}{!}{
\rotatebox{-90}{
\includegraphics{mag_w_10k.ps}
}}
\hspace*{2mm}
\resizebox{0.32\textwidth}{!}{
\rotatebox{-90}{
\includegraphics{r_w_10k.ps}
}}
\hspace*{2mm}
\resizebox{0.32\textwidth}{!}{
\rotatebox{-90}{
\includegraphics{sn_w_10k.ps}
}}
\caption{
An example of the distribution of lensing weights for 10,000 measured galaxies in one 
typical CFHTLenS field,
W1m2m3.  Galaxies were measured on seven exposures to a limiting magnitude $i'<24.7$.  
Weights are shown as a function of ({\em left}) apparent magnitude $i$,
({\em centre}) fitted semi-major axis disk scalelength $r$ and
({\em right}) signal-to-noise ratio $\nu_{\rm SN}$.
\label{fig:weights}
}
\end{figure*}

\subsection{Shear estimation and noise bias}\label{shear}
Papers\,I\,\&\,II discussed at length the advantages of bayesian estimation, but also noted
that a bias would arise in shear estimates if the likelihood distribution for each galaxy
were multiplied by an ellipticity prior before the estimation process.  The existence of
that bias demonstrates that a bayesian method of measuring the {\em ellipticity} of an 
individual galaxy is not in itself 
a complete method for measuring {\em shear} in a sample of galaxies.
A solution proposed
at that time was to calculate the `sensitivity', which measured the biasing effect on a shear
estimator of applying an ellipticity prior.  In Paper\,I, the sensitivity was defined as
$s \equiv \partial\left\langle e_\alpha \right\rangle / \partial g_\alpha$, a calibration of how the
mean ellipticity of a sample $\left\langle e_\alpha \right\rangle$, for ellipticity component $\alpha$,
responds to a cosmological reduced shear $g_\alpha$ in that component.  Cross-terms were ignored.
Analogous approaches to the sensitivity, albeit not in this bayesian context, 
have been discussed by \citet{bernstein02a} and \citet{kaiser00b}.

An immediate difficulty with this approach is that the sensitivity should be a rank\,2 tensor,
but in fact even if only the diagonal terms are considered, the sensitivity is not isotropic,
and application of sensitivity as a correction can lead to an orientation bias in the measured
shear. As the sensitivity is calculated from the measured likelihood surface, and as those
surfaces are biased (Appendix\,\ref{appendixC}), poor results were obtained in CFHTLenS when
this approach was used.

An improved alternative, used for CFHTLenS, is a likelihood-based estimator of ellipticity, in which all
nuisance parameters were marginalised over using a fully bayesian approach as previously
described, but where the ellipticity estimate is based solely on the likelihood distribution
rather than a posterior probability distribution.  It was argued in Paper\,I that such an
estimator should be an unbiased estimator of shear, even though the statistical distribution
of galaxy ellipticities is broadened by the measurement noise:  i.e. the likelihood estimator
has no correction for noise, but should respond linearly to a cosmological shear, when
averaged over many galaxies (when ellipticity is defined as in Section\,\ref{galaxymodels}).
However, the non-linearity of the transformation from a pixel basis to ellipticity does
result in a `noise bias' in shear estimates \citep{refregier12a, melchior12a} which is discussed below.

We should also consider which estimator to use for the ellipticity value of each galaxy.  If
we consider the (simplified) shear measurement process of averaging galaxy ellipticities,
where each galaxy has its own PDF of ellipticity, the PDF of the mean is given by the convolution
of the individual PDFs, suitably rescaled. According to the central limit theorem, the PDF
of the mean should be gaussian, with expectation value given by the mean of the individual galaxy PDF 
means.  Thus to make a fully probabilistic estimator of shear, we should use the mean 
ellipticity as our estimate for each galaxy, not the maximum likelihood estimate.  
Unfortunately, such an estimator is biased,
as discussed in Appendix\,\ref{appendixC}.  For the case where the galaxy size is known precisely,
the maximum likelihood ellipticity should be unbiased and may be a better estimator to use,
but this is not true if the galaxy's size is a free parameter:
there is an inevitable degeneracy in the fits, between size and ellipticity, which
causes a further bias in the likelihood surfaces \citep[see also][]{refregier12a, melchior12a}.
Unlike the case of a pure likelihood likelihood estimator, bayesian marginalisation over the
size parameter mitigates this effect, because small galaxy sizes are downweighted by the size prior,
but it does not eliminate it.  A fully bayesian estimate of ellipticity should control the noise bias,
but currently we do not have a satisfactory method of correcting for the bias on the shear introduced by
the application of an ellipticity prior discussed as the start of this subsection.

To partially address the problem of not being able to use a bayesian estimate for the ellipticity, 
we downweighted
likelihood values at ellipticities where the bayesian posterior probability was low.  
We formed a posterior probability surface
by multiplying by the ellipticity prior, and rejected any posterior probability
values less than some threshold.  The threshold was set to 1\,percent of the maximum
posterior probability.  Applying our prior knowledge of the ellipticity distribution in this way
allowed consistent rejection of the `small, highly elliptical' tail on the likelihood surface 
described above.  The galaxy's
ellipticity was then estimated from the mean of the {\em likelihood} distribution of the ellipticity 
samples that were above the posterior threshold.  
This somewhat {\em ad hoc} approach is not a substitute for a fully bayesian approach, but it did
reduce the residual systematic signals discussed in Section\,\ref{systematics}.
We still expect bias in the estimator at low signal-to-noise ratio, although the lowest
signal-to-noise galaxies also had low weights in the analysis.  
An empirical correction for the
noise bias was derived from the simulations described in Section\,\ref{sims}.  One might hope that,
in future, a better ellipticity estimator could be constructed from the likelihood or bayesian
posterior probability surfaces which is immune to noise bias and which leads to an
unbiased estimate of shear.  To date, we have not achieved such an estimator, but have found an approach
that yields satisfactory results for CFHTLenS.

\subsection{Galaxy weighting}\label{weights}
Faint galaxies have more noisy ellipticity measures than bright galaxies, or, equivalently, have broader
likelihood surfaces. Given that the full likelihood surfaces for the galaxies' ellipticities
are not used in making cosmological shear estimates, 
each galaxy should be weighted according to the width of its likelihood surface in the cosmological analysis. 
We calculated an inverse-variance weight, defined as 
\begin{equation}
w = \left[ \frac{\sigma_{\rm e}^2e_{\rm max}^2}{e_{\rm max}^2 - 2\sigma_{\rm e}^2} + \sigma_{\rm pop}^2 \right]^{-1}
\label{eqn:weight}
\end{equation}
where $\sigma_{\rm e}^2$ is the 1D variance in ellipticity of the likelihood surface,
$\sigma_{\rm pop}^2$ is the 1D variance of the distribution of ellipticity of the galaxy population
and $e_{\rm max}$ is the maximum allowed ellipticity, assumed here to be the maximum disk ellipticity
as in equation\,\ref{eqn:edisk}.  In the limit where $e_{\rm max}\rightarrow\infty$, this definition of the weight tends towards a conventional form, $w \rightarrow (\sigma_{\rm e}^2 + \sigma_{\rm pop}^2)^{-1}$.  
A finite value of $e_{\rm max}$ was included to ensure that a 
galaxy with a likelihood surface which is uniform for $e < e_{\rm max}$, 
and which therefore conveys no information about the galaxy shape, has zero weight.
For a gaussian likelihood surface and prior (and in the limit $e_{\rm max}\rightarrow \infty$), $w$
has the same value as the sensitivity (Section\,\ref{shear}), except for its normalisation
(Paper\,I, Section 2.5).  
The expressions differ in the (actual) case of non-gaussian probability
surfaces, but $w$ has the advantage that it is not by definition anisotropic.  In the limit
of low-noise measurements, the weight becomes dominated by the `shape noise' of the galaxy
population, as expected for weak-lensing studies \citep{bernstein02a}. Fig.\,\ref{fig:weights}
shows an example of the distribution of weights in one of the CFHTLenS fields, as a function
of $i'$-band magnitude, fitted galaxy major-axis scalelength, and signal-to-noise ratio, $\nu_{\rm SN}$.  
The weight falls off sharply for $\nu_{\rm SN} \la 20$.  The maximum value of the weight
is determined by the value of $\sigma_{\rm pop}=0.255$, obtained from the ellipticity prior
described in Appendix\,\ref{appendixB}.

However, although the weight is designed to be isotropic, it is possible for there to exist a 
form of `orientation bias' similar to the selection bias discussed by \citet{kaiser00b} and
\citet{bernstein02a}.  In the case of the selection bias, we are concerned about galaxies
being omitted from the sample if they are cross-aligned with the PSF.  However, 
even for detected galaxies, the weight also
may be larger for galaxies that are aligned with the PSF than for those that are cross-aligned,
because the likelihood surface may be more sharply peaked.
We did not make any attempt to correct explicitly for this effect, but it may manifest itself
as a correlation between weighted galaxy shapes and the PSF, as shown in Section\,\ref{simanalysis}.

\subsection{The treatment of large, blended or complex galaxies \label{sec:blends}}
The galaxy models were fitted to each galaxy in subregions extracted around each galaxy 
(`postage stamps') of size 48 pixels (approximately $9''$), this choice being a compromise between having a 
large enough region to correctly model the galaxies but small enough that the compute time
for operations such as the fast Fourier transform was not prohibitively slow.  If each galaxy
had $n$ useful exposures, then $n$ postage stamps were created and models jointly fitted, allowing
for astrometric offsets and camera distortion as described in 
Sections\,\ref{image_combination}\,\&\,\ref{distortion} below.
Inevitably, some galaxies had sizes too large to be fitted in this size of postage stamp; such galaxies
were excluded from the analysis.

In some cases, two or more neighbouring galaxies appeared within the same postage stamp.  The
algorithm can only fit one galaxy at a time, so the solution adopted was to first see whether it
was possible to mask out one galaxy (set its pixel values equal to the background) without
disturbing the isophotes of the galaxy being fitted.  To this end, a co-added image postage stamp was
created, averaging all the exposures available for that galaxy, shifted so the relative
positions agreed to the nearest pixel, which was then smoothed by a gaussian of 
FWHM equal to that of
the local PSF.  Isophotes were created for each smoothed galaxy: if a separate galaxy or other
object was identified with non-touching isophotes, at a level of twice the smoothed pixel noise, 
that other galaxy was masked out and the fitting
would proceed.  Such close pairs of galaxies are thus included in the output catalogues from
CFHTLenS. We note, however, that low-level light leaking below the two-sigma isophote could still
contaminate the measurement, and thus we expect the ellipticity measurements of galaxies in close pairs,
whose isophotes may be contaminated by their neighbour, to be artificially correlated.

Within each postage stamp, it may be that some pixels should be masked because of image defects.  The
{\sc theli} pipeline provided images of pixel masks to be applied.  If such masked pixels occurred within the
two-sigma isophote of a galaxy on one individual exposure, that exposure was not used in the joint
analysis.  If such pixels occurred outside the two-sigma isophote, the pixel values were set equal to
the background and that masked exposure was used in the joint fitting.

\begin{figure}
\centering
\resizebox{0.23\textwidth}{!}{
\rotatebox{0}{
\includegraphics{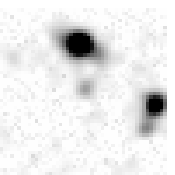}
}}
\resizebox{0.23\textwidth}{!}{
\rotatebox{0}{
\includegraphics{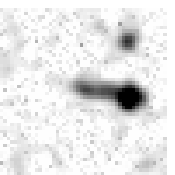}
}}

\vspace*{3mm}

\resizebox{0.23\textwidth}{!}{
\rotatebox{0}{
\includegraphics{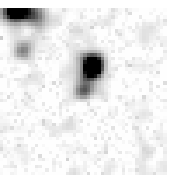}
}}
\resizebox{0.23\textwidth}{!}{
\rotatebox{0}{
\includegraphics{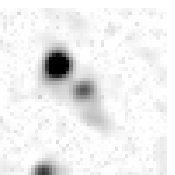}
}}

\caption{
Examples of four galaxies excluded from measurement by the criteria described in Section\,\ref{sec:blends}, in field W1m0m1.  Each panel shows a coadded image 48 pixels (approximately $9''$) square, centred on each target galaxy, and the inverted grey scale is linear up to some maximum value which varies between images.
\label{fig:blends}
}
\end{figure}

Other galaxies may be sufficiently close that their smoothed isophotes overlapped, and there may also be
individual galaxies with complex morphology, not well described by a simple bulge-plus-disk model.
These galaxies were identified using a deblending algorithm, testing for the presence of significant
independent maxima in the smoothed surface brightness distribution\footnote{
The algorithm was similar to that of \citet{beard90a}.  Maxima in the
smoothed surface brightness distribution associated with the target
galaxy were identified, and regions `grown' around those maxima by
successively lowering a threshold isophote level from that maximum
level. Pixels above the threshold were either identified with the
corresponding maximum of any identified pixels that they touched, or
otherwise were defined to be a new, secondary, maximum.
Regions with fewer than 8 pixels were amalgamated into any touching
neighbours.  If multiple regions remained after this process, within
the limiting $2\sigma$ isophote, the galaxy was flagged as `complex'.}.  
Any such complex or blended galaxies that
were found were excluded from the analysis.  
A further criterion was imposed, that the intensity-weighted centroid of a galaxy, measured from the
pixels within the smoothed $2\sigma$ isophote, should lie within 4\,pixels
of the nominal target position: this criterion guarded against any blended galaxies
that had been identified as blends in the original input catalogue but 
that had not been identified by the other tests described in this section.
Some examples of images of galaxies excluded by these criteria are shown in Fig.\,\ref{fig:blends}, which shows examples of the stacked, smoothed images
used for testing for object complexity.  Visual inspection indicated that the great majority of galaxies 
excluded in this way had isophotes that overlapped with neighbouring galaxies.

The fraction of galaxies that were excluded in this way varied somewhat between fields, as the criteria
were affected by the size of the PSF.  Typically, 20\% of galaxies were excluded.  Although this
fraction seems high, such a loss of galaxy numbers does not significantly degrade the signal-to-noise
of the final cosmological analysis, but it does help ensure that galaxies whose measurements would be
poor because of their size, or because they would be poorly modelled, have been excluded.  These exclusion
criteria are likely to introduce small-scale selection effects into the galaxy distribution (e.g. 
neighbouring galaxies would have been classed as being blended with greater or lesser probability depending on
how they were aligned with respect to the PSF) and so lensing signals on arcsec scales, $\la 5''$, should
be excluded from analyses of this survey, even though nominal measurements are reported in the
output catalogues.  We note that the exclusion of some fraction of close pairs of galaxies may
introduce a bias at a level of a few percent into cosmological parameters \citep{hartlap11a}: we
do not currently have any way to estimate the size of this bias in an actual survey such as CFHTLenS,
without a detailed model of the true distribution of galaxy pairs and of the effect of the measurement
process on those pairs.

\section{Optimal combination of multiple images}\label{image_combination}
The algorithm presented in Papers\,I \& II, and also the
simulations of the GREAT08 \citep{bridle10a} and GREAT10 \citep{GREAT10} challenges,
assume that each galaxy is measured on a single image.  However, actual galaxy
surveys use combinations of multiple exposures in the same waveband, or even across different filters.  
The reasons for having multiple exposures in the same filter are: 
(i) to increase the dynamic range of the observations; 
(ii) to prevent an
excessive build-up of 
cosmic ray artifacts on any one image; 
(iii) to allow dithering of observations, filling in
gaps where CCD boundaries or CCD artifacts prevent useful data being obtained and 
mitigating the effects of the finite pixel sampling.  Thus any shear
measurement method should make optimal use of such multiple images. In CFHTLenS typically
seven dithered exposures were obtained in each field (Section\,\ref{sec:cfhtlens}).

It is common practice for multiple images to be averaged together to make a single image
on which shear measurement is performed.  This is problematic for high-precision 
galaxy shape measurement, for several reasons.
\begin{enumerate}
\item In order to make the averaged image, the individual images need to be first
interpolated onto a common pixel grid.  The interpolation causes distortion of the PSF,
but because of the slightly differing local plate scale between exposures,
the interpolation kernel varies across the image on scales typically less than
arcminutes, leading to patterns analogous to Moir\'{e} fringes and 
introducing artificial PSF variations on arcminute scales. These
are difficult to calibrate from star measurements owing to the finite density of
stars in typical observations.  Knowing the interpolating kernel, the effect of the 
interpolation could in principle be corrected for at the location of each galaxy, but
this is not a straightforward process.
\item The interpolation introduces correlation between pixels into the noise, which is
also problematic to correctly allow for in shear measurement methods. That noise
correlation also varies on arcminute scales.
\item Usually the PSF differs between each observation, so that averaging
does not represent an optimal combination of images.
\item The co-addition of PSFs of differing shapes and orientations may lead to a complex stacked PSF which might be more problematic to measure and model, depending on the method adopted.
\end{enumerate}
A likelihood-based method provides a straight-forward way of overcoming all four issues. The galaxies on each individual image can be fitted by models that have been convolved with the correct PSF for that image, and the likelihoods simply multiplied to obtain the final combined likelihood.  Images of poor quality, perhaps with a worse PSF, have a broader distribution of likelihood values than those of better quality, and hence the combined likelihood represents an optimal combination of information.  Because models are fitted to each individual image, there is no need for interpolation, avoiding the first two problems listed above.  Tests of the {\em lens}fit algorithm on stacked images, not taking into account the above effects, produced significantly worse results, with large cross-correlation signals between galaxies and the PSF, than were obtained from the joint analysis of individual exposures presented here. 

In this method, the likelihoods are calculated on each image as a function of all the model parameters, including the unknown position of a galaxy. Optimum results are obtained by assuming that the galaxy's position should be the same, in celestial coordinates, on each image - provided the relative registration of the images is known, the positions on each exposure should be the same and should not be treated as independent model parameters.
This approach requires highly accurate
knowledge of the relative astrometric registration of the images, however, otherwise an artificial shear would be introduced which would be highly correlated between neighbouring galaxies.  A similar problem arises when averaging images into a single stacked image: astrometric errors degrade the PSF.  However, in that case, at least the PSF then is measured from the stacked image, so the degraded PSF is, in principle, still correctly measured (unless astrometric errors vary on small length-scales across the image).  In the case of the likelihood-based method, we do not have this luxury, because the PSFs are measured on individual frames.  Thus, it is essential to have an accurate measurement of the relative image registrations.  Those registration shifts can then be applied to each of the models fitted to the individual images. This process must be accurate to significantly less than one pixel, since astrometric errors are correlated over large scales across an image, and if uncorrected, would introduce an apparent shear.

The way these relative astrometric registrations were measured for
CFHTLenS was, for each image, to cross-correlate the local PSF model
with each star, thus measuring, from the maximum of the cross-correlation, 
the apparent location of each star with
respect to its nominal celestial coordinate with maximum precision.  A
low-order polynomial function was then fitted to those registration
shifts across each CCD, and the shift function applied to the positions of
the PSF-convolved galaxy models when fitting each exposure of each galaxy.
A further sub-pixel correction was applied to allow for the fact that an individual galaxy's
postage stamp was extracted from each exposure based on the nominal celestial
coordinate (using the same coordinate transformation as for the stars used to create the PSF model),
but was centred at an integer pixel value (i.e. no interpolation of data was
done when extracting each postage stamp).
The final shifts applied were always
sub-pixel, and usually were less than one-tenth of a pixel.

\begin{figure}
\centering
\resizebox{0.15\textwidth}{!}{
\rotatebox{0}{
\includegraphics{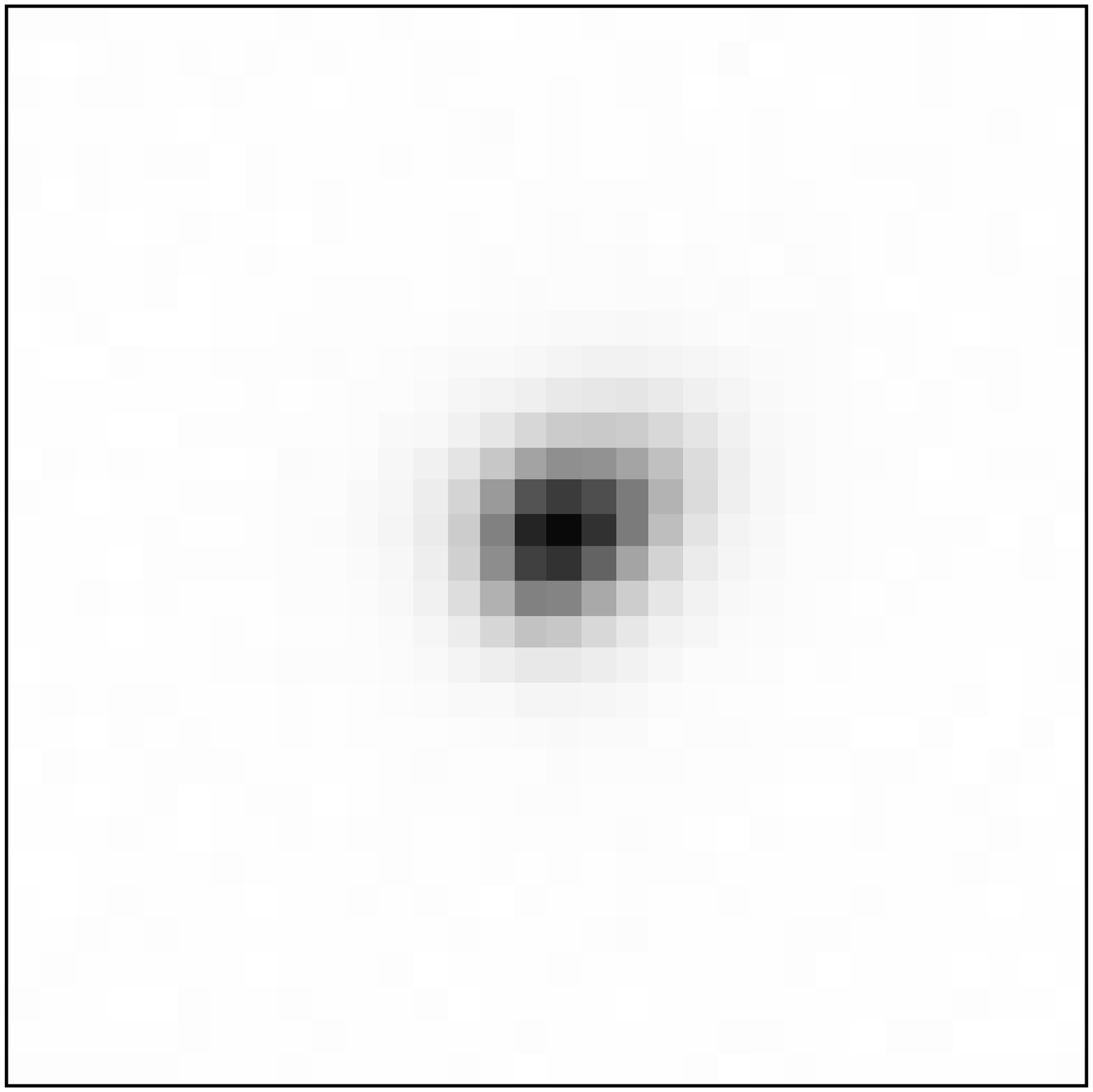}
}}
\resizebox{0.15\textwidth}{!}{
\rotatebox{0}{
\includegraphics{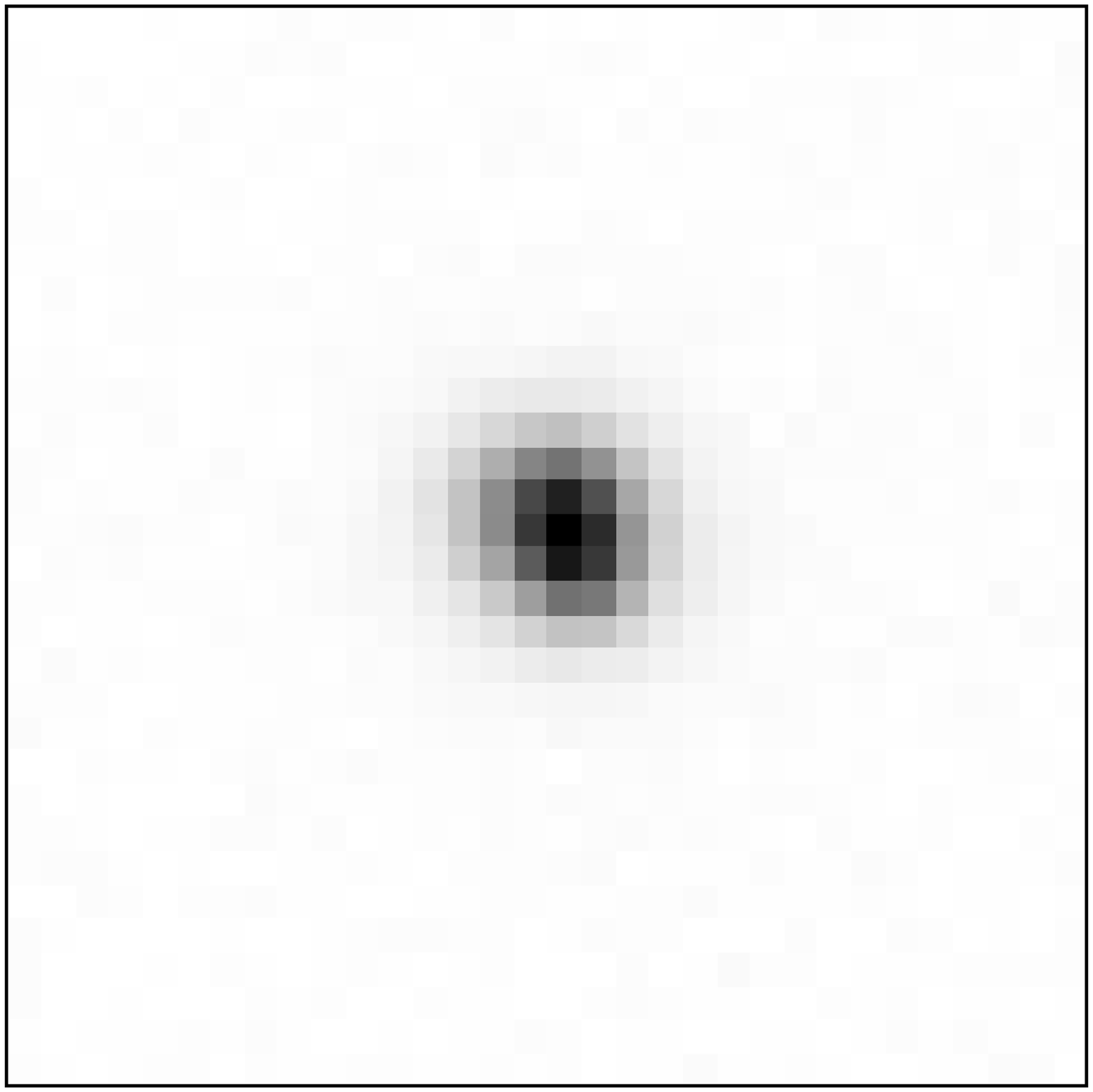}
}}
\resizebox{0.15\textwidth}{!}{
\rotatebox{0}{
\includegraphics{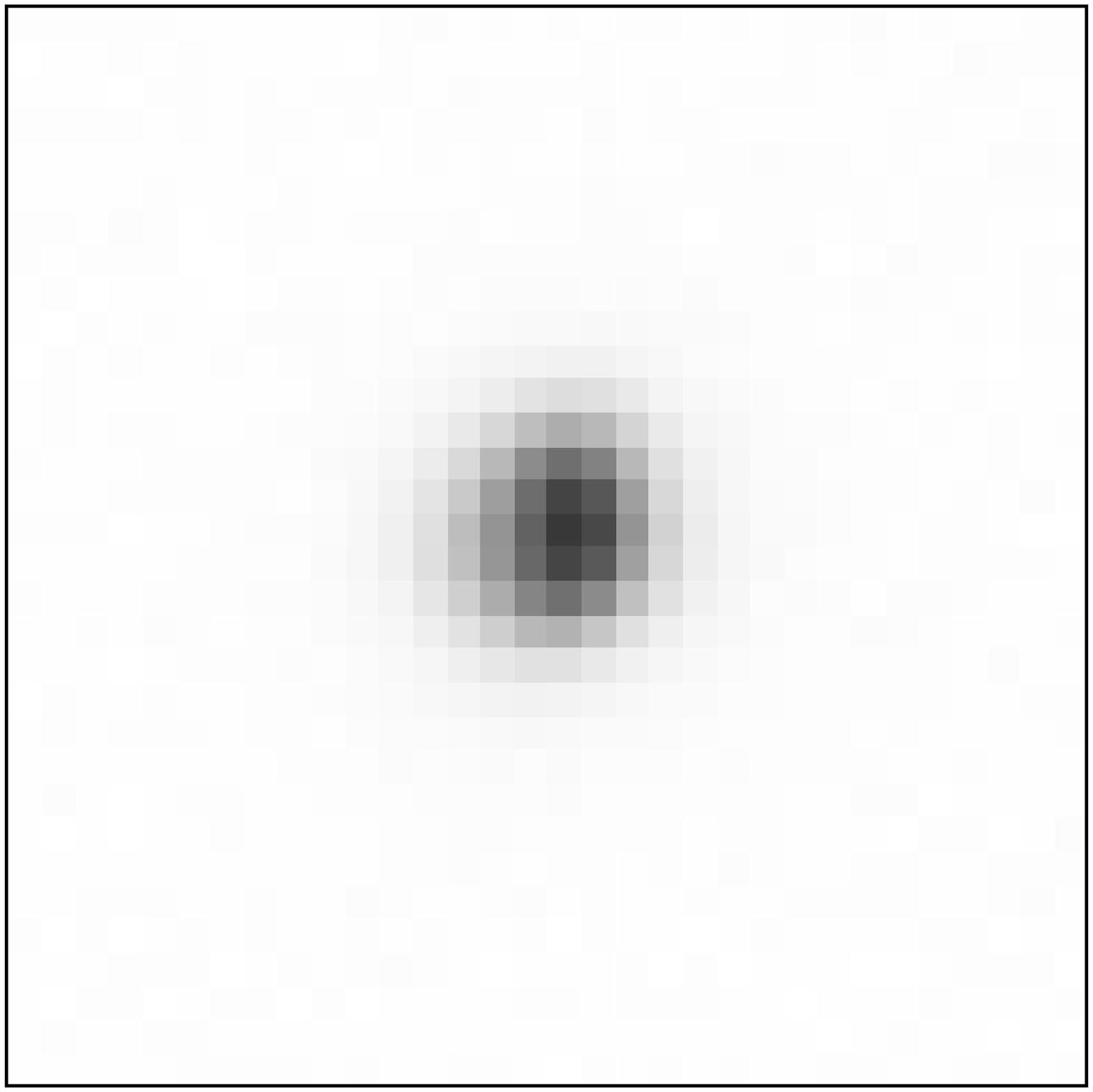}
}}

\resizebox{0.15\textwidth}{!}{
\rotatebox{0}{
\includegraphics{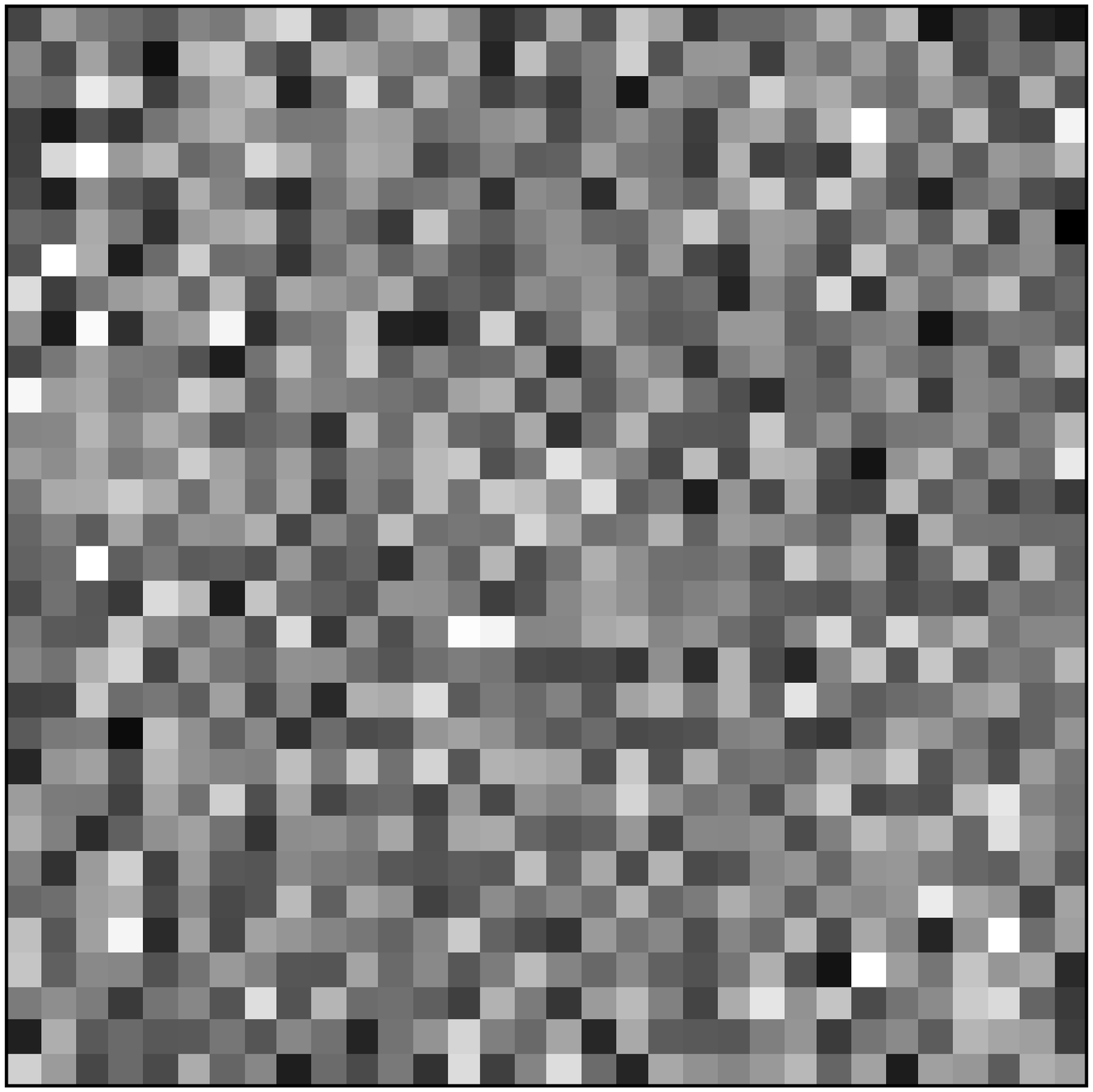}
}}
\resizebox{0.15\textwidth}{!}{
\rotatebox{0}{
\includegraphics{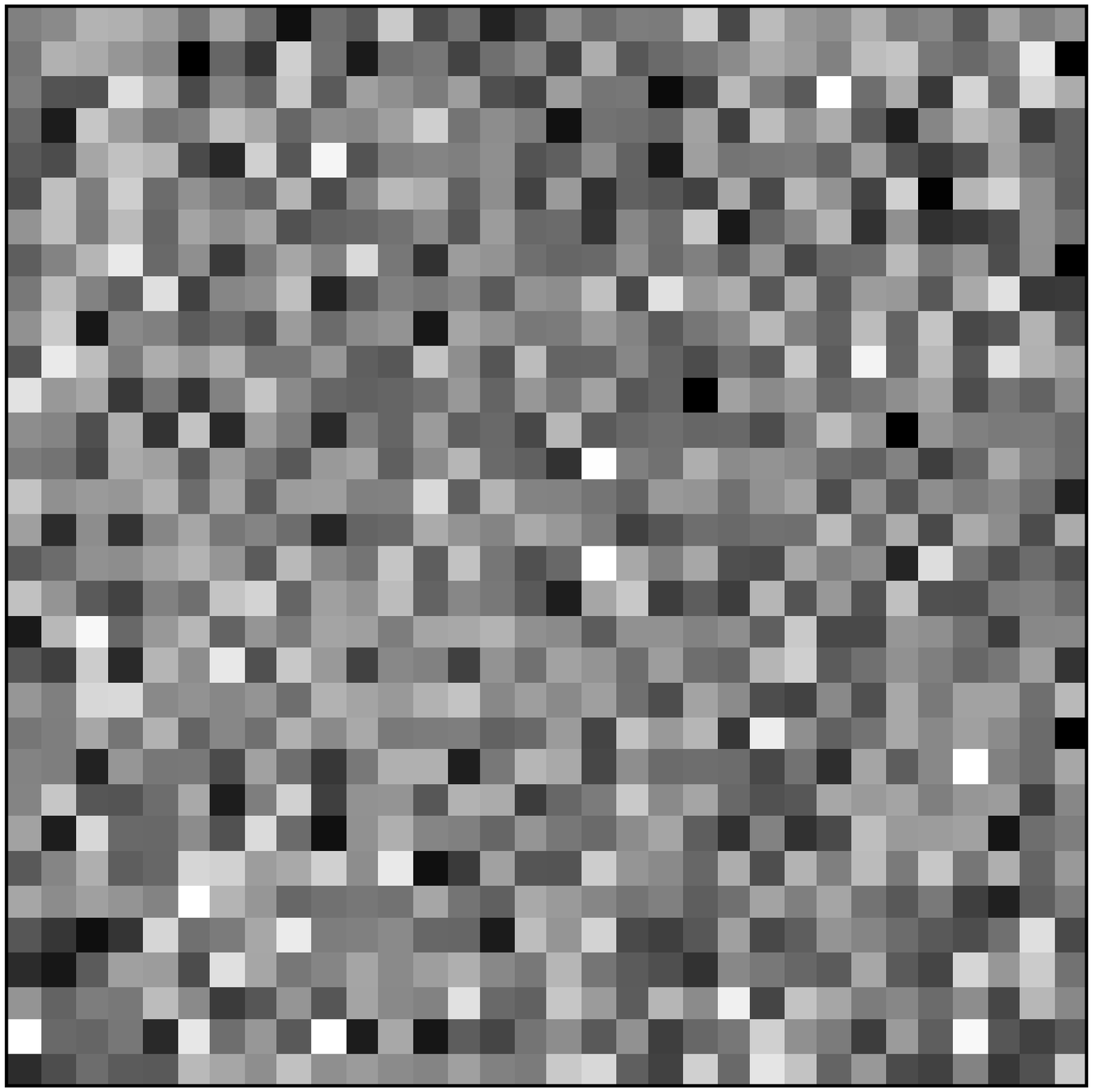}
}}
\resizebox{0.15\textwidth}{!}{
\rotatebox{0}{
\includegraphics{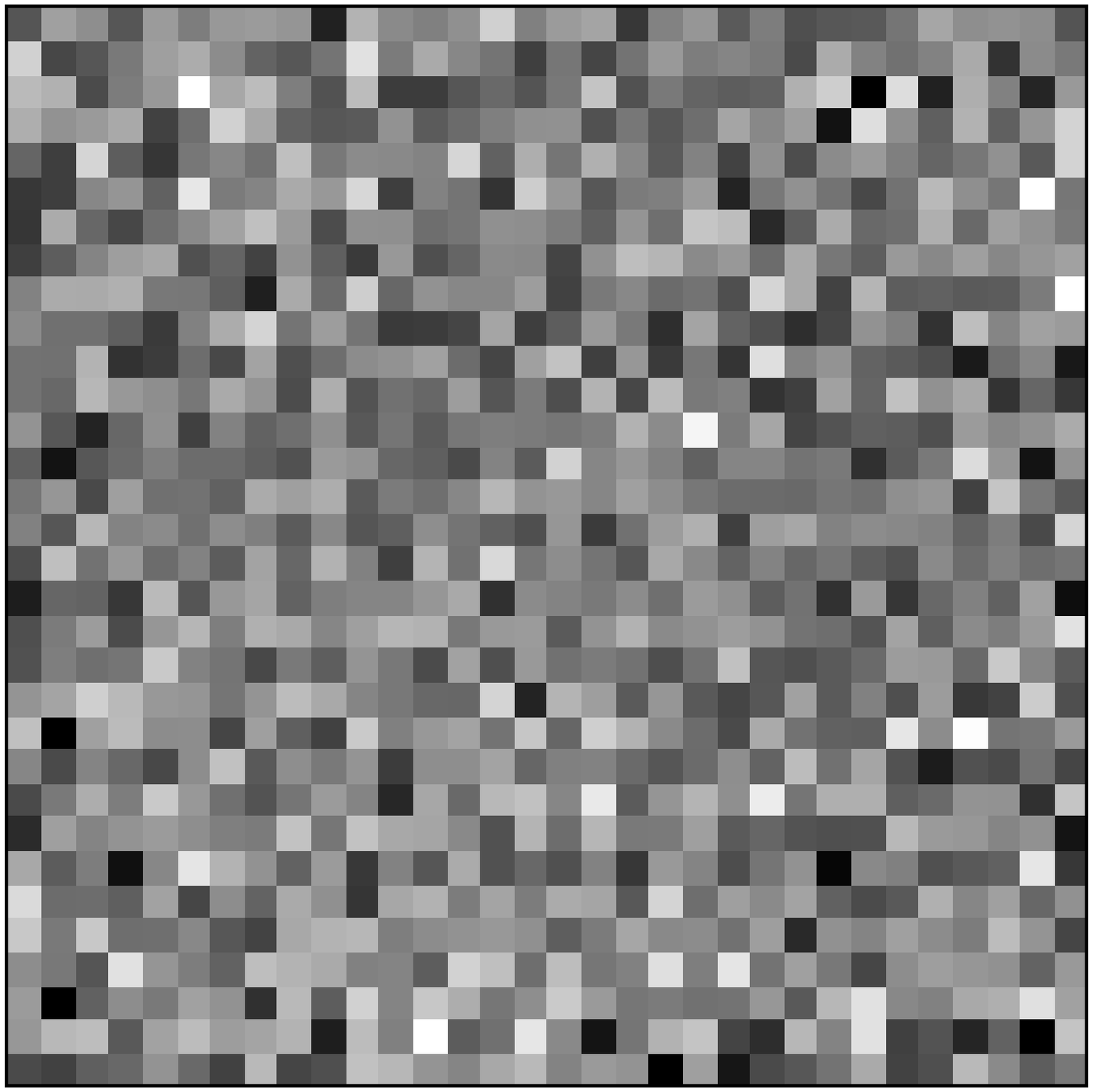}
}}

\resizebox{0.15\textwidth}{!}{
\rotatebox{0}{
\includegraphics{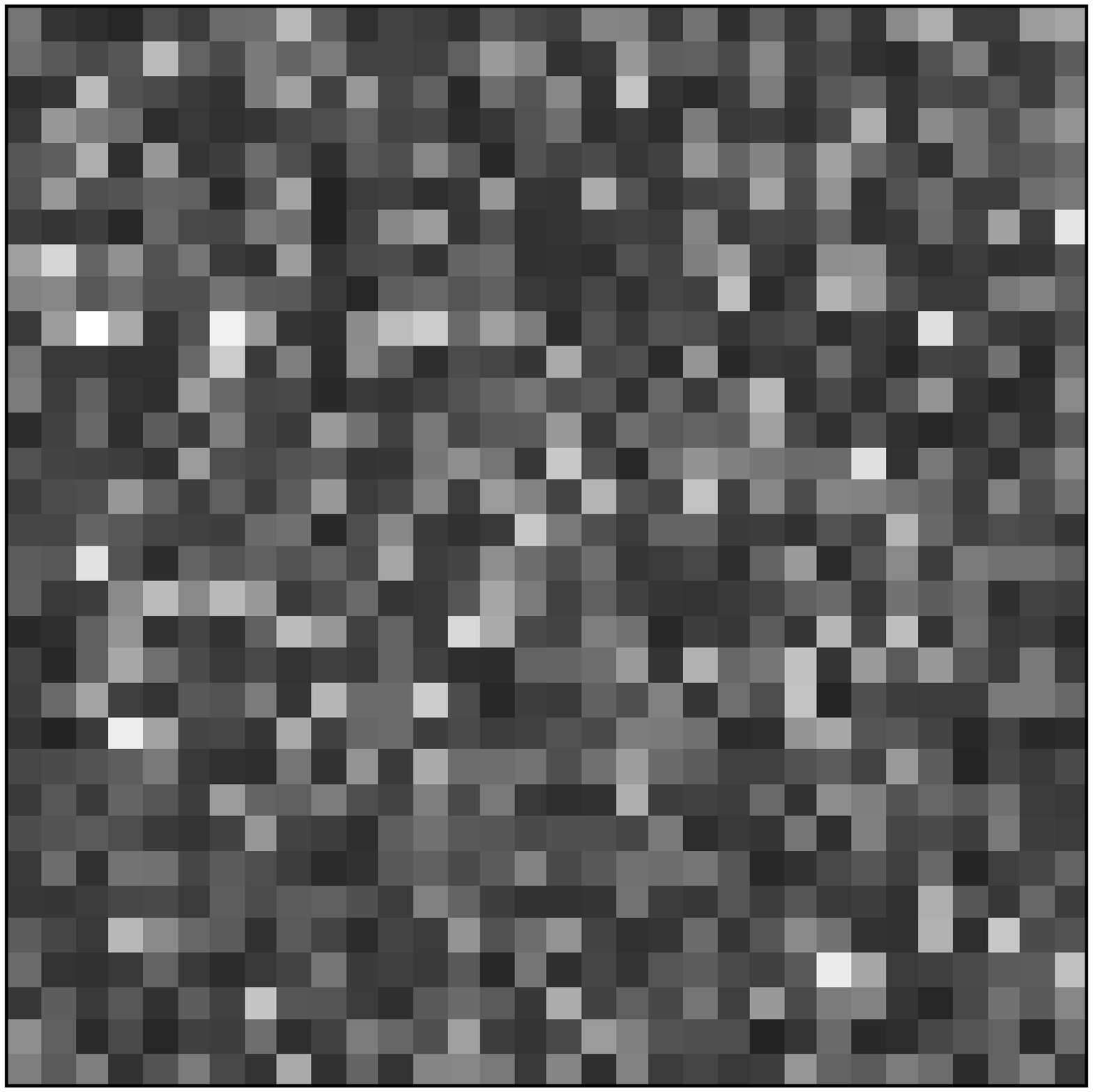}
}}
\resizebox{0.15\textwidth}{!}{
\rotatebox{0}{
\includegraphics{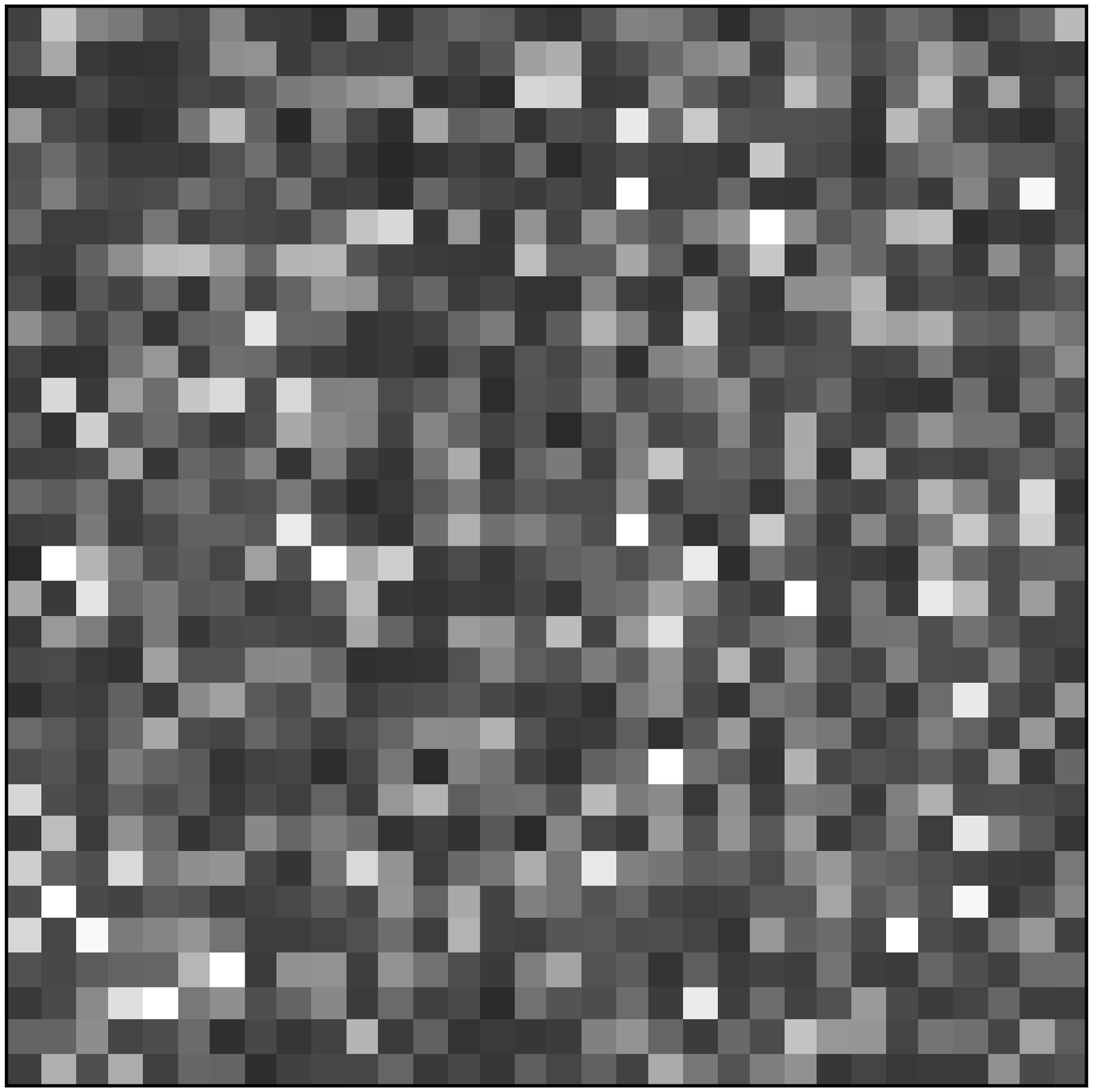}
}}
\resizebox{0.15\textwidth}{!}{
\rotatebox{0}{
\includegraphics{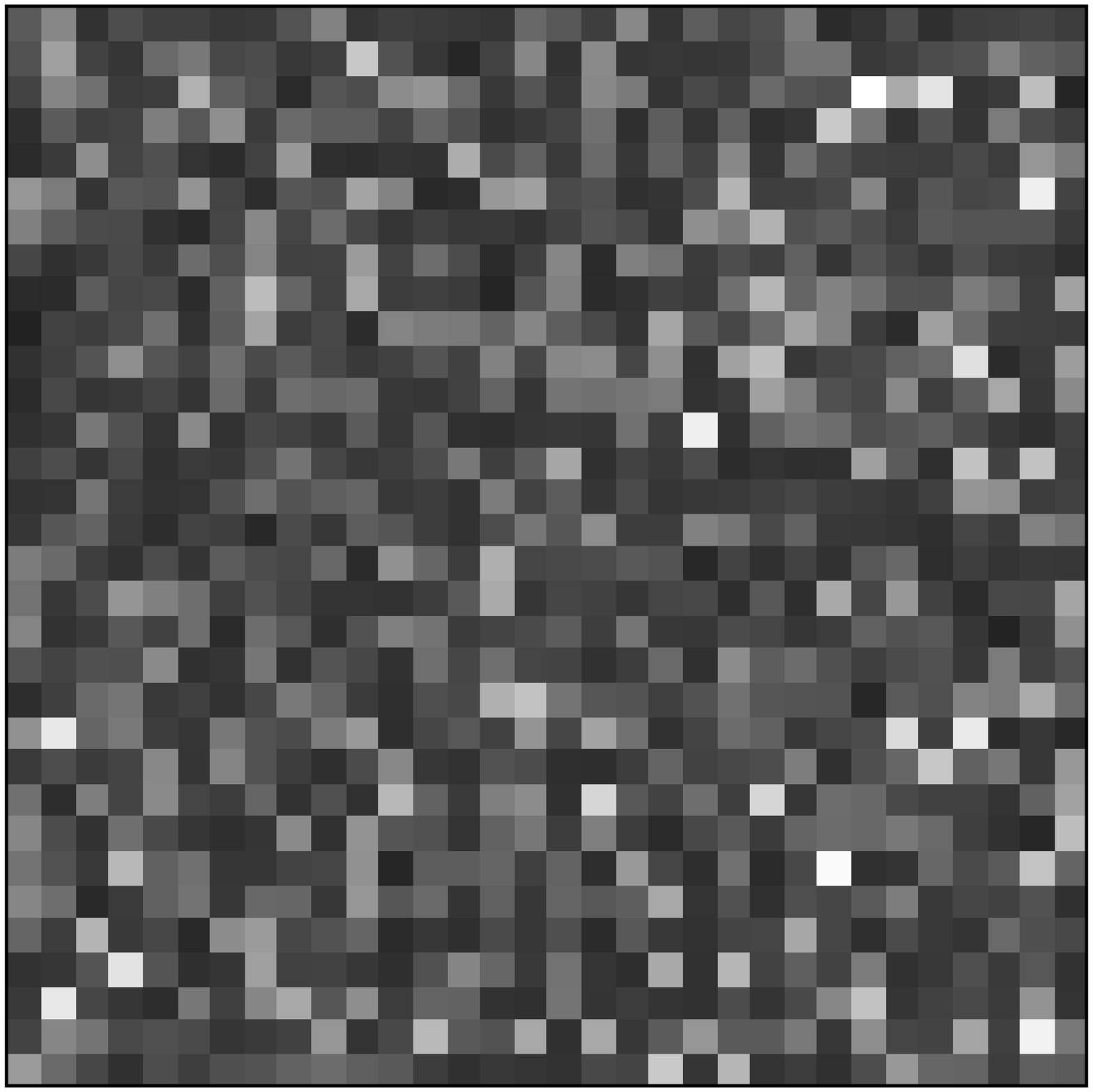}
}}
\caption{
Example PSF models derived in a typical CFHTLenS exposure: ID 720093 in field W1p3m0.  
The upper panels show the model PSF sampled at three locations across the MegaCam field, 
with a linear greyscale varying from 0 to 0.05, where the
values are relative to the total flux in the PSF.  Each panel is 32 pixels (approximately $6''$) square.
The centre panels show the mean residuals, averaged over all the stars selected on that CCD,
between the PSF model and the star images, with a linear greyscale varying over the range
$\pm 5\times 10^{-4}$.  No residuals are larger than $7\times 10^{-4}$ anywhere in these panels.
The lower panels show the rms variation in the model, measured as described in the text, with a 
linear greyscale over the range 0\,(black) to $5\times 10^{-4}$\,(white), 
with maximum rms $5.6 \times 10^{-4}$.
The PSFs and the residuals are shown:
(left) at the centre of the CCD in the highest right ascension, lowest declination part of the MegaCam field;
(centre) at the centre of the CCD immediately below the centre of the MegaCam field;
(right) at the centre of the CCD in the lowest right ascension, highest declination part of the MegaCam field;
\label{fig:psfimages}
}
\end{figure}

Similarly to the issue of celestial position of a galaxy, optimum joint analysis of multiple exposures should
also assume that a galaxy has an invariant, but unknown, true flux, and that the relative measured flux on each 
exposure is related to the true flux through a known calibration relation.  As the photometric calibration was
carried out over many bright stars, we assumed the uncertainty in the photometric calibration was negligible compared
with the photometric uncertainty associated with each image of an individual galaxy.  Thus a galaxy's flux
was treated as a single unknown parameter to be marginalised over, using exposures that had the photometric
calibration applied to them.  

In  principle, the same methodology could include exposures obtained in other
filters, provided the filters are sufficiently close in wavelength that the same model provides a 
good description of each galaxy's structure at all the observed wavelengths. 
In that case, the galaxy's flux and bulge fraction values 
in each filter should be added as additional free parameters to be marginalised over.
In CFHTLenS, only the deep, good-seeing $i'$-band exposures were used.

\section{Point spread function estimation}\label{sec:psf}
As in other lensing methods, we measured the PSF from images of stars on the same exposures as the galaxies being
measured. The PSF was represented as a set of pixels at the same resolution as the data, and in the model-generation process, galaxy models were convolved with that PSF model.  The PSF varied over the field of view: in MegaCam it varies both over the full optical field and also between the individual CCDs that comprise the full camera, and discontinuities in PSF are detectable between CCDs.

\begin{figure*}
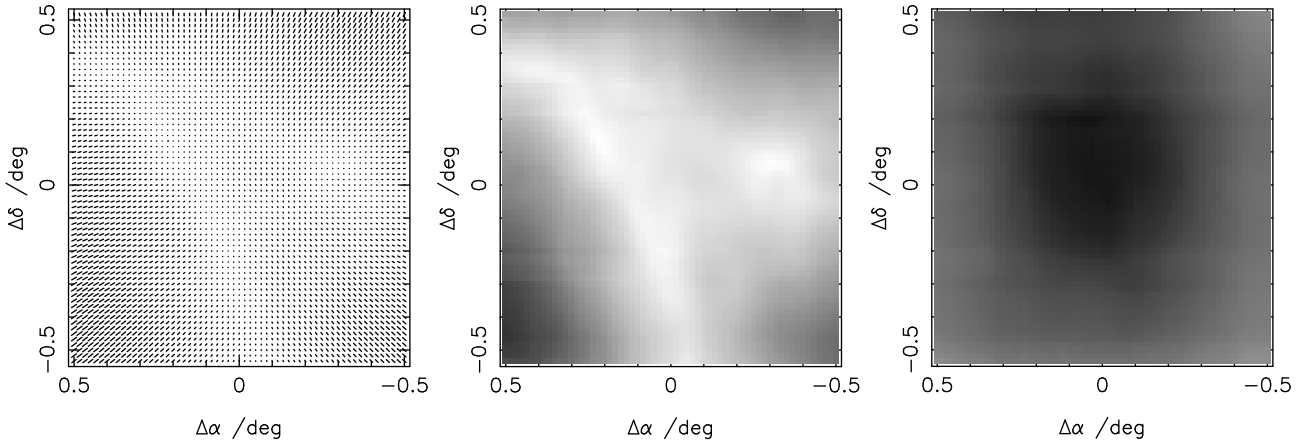

\resizebox{0.32\textwidth}{!}{
\rotatebox{-90}{
\includegraphics{w1m1p2_sticks_v7.ps}
}}
\resizebox{0.32\textwidth}{!}{
\rotatebox{-90}{
\includegraphics{w1m1p2_grey_v7.ps}
}}
\resizebox{0.32\textwidth}{!}{
\rotatebox{-90}{
\includegraphics{w1m1p2_strehl_v7.ps}
}}
\caption{The variation of measured PSF over one of the CFHTLenS fields with a highly
variable PSF, W1m1p2.  
The plot has been made by combining information from seven exposures, which are somewhat
offset with respect to each other.  PSF ellipticities have been simply averaged.
{\em (left)} stick plot showing the variation of measured PSF ellipticity across the field.  
The maximum stick length on the plot corresponds to ellipticity of 0.13.
{\em (centre)} greyscale showing the ellipticity amplitude, with black having an ellipticity of 0.16. 
{\em (right)} greyscale showing the fraction of the light in the central pixel of the PSF, with black
indicating a fraction 0.1.}
\label{fig:psf}
\end{figure*}

To generate a pixel representation of the PSF, an initial catalogue of candidate stars was selected from a stacked image of all exposures.  Stars were selected using a conventional cut in the apparent size-magnitude plane, identifying by eye the locus of stars in the size-magnitude plane and selecting a box around that locus.
The selection was further refined by applying a three-waveband $gri$ selection.  First, a stellar locus was defined in colour space by measuring the colour-space density of the input catalogue and rejecting any object which lay in a low-density region.  Images of a subset of the resulting catalogues were inspected by eye and found to comprise clean samples of stars (note that this procedure was relevant only for the generation of bright star catalogues
for PSF creation and was not used for the classification of faint objects in the lensing measurement).
More details of the generation of the PSF star catalogues are given by \citet{heymans12a}.

A pixelised PSF model was then created for each exposure by first centering each star according to a measurement of its centroid, then fitting each recentred pixel value as a two-dimensional polynomial function of position in the telescope field of view (the centering process is discussed below).
For the interpolation in CFHTLenS, a third-order polynomial was found to produce good results on a random subset of the survey, with no improvement found using fourth-order, where the test used was to measure the residual cross-correlation between the PSF model and galaxy ellipticities 
(Section\,\ref{systematics}).
To allow for the discontinuities in the PSF across the boundaries between CCDs, some coefficients of the polynomial could be allowed to vary between CCDs, with the remaining coefficients being fixed across the field of view. Good results were found allowing the constant and linear terms to vary in this way.

The polynomial fits were constrained by the images of the stars in the input catalogue, where only stars with integrated signal-to-noise ratio larger than 20 were used, and where each star was weighted by the function
$w_s = \nu_{\rm SN, star}^2/(\nu_{\rm SN, star}^2 + 50^2)$,
where $\nu_{\rm SN, star}$ was the star's integrated signal-to-noise ratio on that exposure and where the
extra term in the denominator prevents any one very bright star from dominating the polynomial fit.  Any star whose central pixel was more than 50\% of the full-well limit for the CCD was not used, and any star which had either pixels that were flagged as not usable 
or pixels that appeared to be part of another object within a radius of eight pixels of the star centroid were also excluded.  Pixels were deemed not usable either if they appeared to
have been affected by a cosmic ray impact, if they were identified as being defective in a `bad pixel' mask, or if they appeared to be affected by charge transfer from saturated stars, in the {\sc theli} pipeline
(Section\,\ref{sec:cfhtlens}). Any individual CCD images that had fewer than 40 usable stars were not included in the analysis, as it was found that smaller numbers of stars could produce PSF models that were not robust to a test of varying the stars selected.

Star centroids were measured iteratively by cross-correlating with the PSF model and finding the centroid location.  Star images were then interpolated so as to be centred on the pixel grid, using sinc-function interpolation, and thus assuming the PSF to be a Fourier band-limited function (i.e. assuming there are no modes above the Nyquist limit of the data).  As this assumption is unlikely to be valid, this is one of the main areas to be addressed in future improvements of the algorithm (see Section\,\ref{future}).

The top panels of Fig.\,\ref{fig:psfimages} show an example of the PSF
pixel model created in one of the CFHTLenS exposures, sampled at three
locations across the MegaCam field (two of the corners and near the
centre).  For comparison, the centre panels of the figure also show
the mean difference between the model and the stars, averaged over all
the stars on each CCD (80, 71 and 105 stars in each of the three
CCDs shown).
The lower panels show the rms variation in the
model, measured by creating 64 bootstrap resamplings of the input star
catalogue, recalculating the entire PSF model for each resampling, and
measuring the rms variation in those mean residuals between the
bootstrap-generated models and the stars from the original catalogue.
The PSF model was fitted globally to all 36 MegaCam CCDs
simultaneously, and the values shown have been evaluated by averaging
over all stars on one CCD.  The images have been normalised so that the sum of
values in the PSF is unity.  Within the central region of the PSF the
typical rms is $\sim 3 \sim 10^{-4}$.  This test shows the
reproducibility of the PSF model generation on the scale of each CCD,
$6.3' \times 14.3'$.

Fig.\,\ref{fig:psf} shows an example of the PSF variation across one
of the CFHTLenS fields.  The quantities plotted are the PSF
ellipticity (orientation and amplitude) and the fraction of light in
the central pixel.  These quantities have been averaged over 7
exposures.  As well as a large-scale radial distortion, jumps in PSF
are visible at CCD boundaries, especially the horizontal boundaries
where there are two larger gaps between detectors, and where those
boundaries appear ill-defined on the figure because of dithering of the 7
exposures.

\section{Distortion correction}\label{distortion}
Wide-field cameras introduce distortion into the field-of-view which can mimic the cosmological shear signal, and this therefore needs to be corrected for.  Because field distortion may vary with telescope position and temperature, and because exposures are dithered, it should ideally be mapped and corrected on each individual exposure.  The distortion information is implicitly
available in the astrometric calibration of the survey: provided the astrometric solution has good absolute accuracy, the distortion can be measured simply by measuring the relationship between pixel and celestial coordinates as a function of position across the field.  The centroid positions of stars, on which the astrometric solution is based, can generally be measured to subpixel accuracy, whereas the scalelengths of distortion variations are typically arcminutes or greater, so that the typical accuracy in distortion correction by this method is shear error of order $10^{-3}$ or better.  

Having measured the distortion on an image, this information was incorporated into the galaxy models by calculating the local linearised distortion transformation and applying that as an affine transformation to the models being fitted to each galaxy on each individual exposure.  In other words, the effect of distortion was corrected by including it in the forward modelling of 
each galaxy on each exposure, 
rather than by trying to remove it from the data.
Because the PSF models had not had distortion removed from them (the stars were not distortion-corrected on input to the models), the distortion was applied to the galaxy models prior to PSF convolution, to ensure
that the distortion and convolution corrections to the models correctly mirrored these effects in the data.
In practice, the models were fitted in Fourier space, so the affine transformation was also applied in Fourier space, using a simple matrix operation to convert the affine transformation between real and Fourier space \citep{bracewell93a}. 
The affine transformation allowed for scale, rotation and shear distortion, and ensured that the ellipticities of all galaxies were measured with respect to the celestial coordinate system, and not any arbitrary pixel grid (i.e. galaxy position angle was defined with respect to the local north-south, east-west axes at each galaxy).   When calculating shear correlation functions or power
spectra from these data on large scales, it may be necessary to correct for the translation of the celestial coordinate
system around the celestial sphere (i.e. there needs to be a `course angle' correction to the differences in position angles
of pairs of galaxies separated on large angular scales, when 
not working on the celestial equator).

\section{Tests for systematic errors}\label{systematics}
We postpone detailed analyses for systematic biases to a companion paper \citep{heymans12a}, however
in this section we discuss the chief effects which may bias the shear measurement results and illustrate the
tests that we have carried out, showing specific examples from individual CFHTLenS fields.

There are a number of possible effects which could lead to the creation of systematic biases in the measured shear.
\begin{enumerate}
\item Inaccurate PSF modelling.
\item Use of galaxy models and prior distributions for the model parameters that are not representative of the true population.
\item Bias in the shear estimator (such as that described in Appendix\,\ref{appendixC}).
\item Inaccurate relative astrometry.
\end{enumerate}
PSF modelling errors could be investigated by comparing the models with the stars used to generate the models (i.e.
a quantitative assessment such as shown in Fig.\,\ref{fig:psfimages}): however, it is difficult to assess how large a deviation
between model and reality can be tolerated, what is really needed is an assessment on the effect of the final
measured galaxy shear.  Such an analysis has been presented by \citet{paulinhenriksson08a, paulinhenriksson09a},
giving a good qualitative picture of the effects expected from PSF modelling errors.  
It is difficult to make a quantitative prediction of the effect on shear without repeating the analysis
with realistic galaxy sizes and shapes, PSF surface brightness distributions that match the CFHTLenS PSFs
and likely PSF modelling errors.  However, the results of \citet{paulinhenriksson08a} show that we expect
modelling errors to result in significant cross-correlation between the PSF model and the galaxy ellipticities, and
furthermore that the cross-correlation is expected to be a strong function of galaxy size.  The use of
incorrect galaxy models or prior distributions is likewise expected to produce size-dependent PSF-galaxy
cross-correlation.  Bias in the shear estimator is expected to produce PSF-galaxy cross-correlation 
with a strong signal-to-noise dependence (Fig.\,\ref{fig:bias}).  

\begin{figure*}
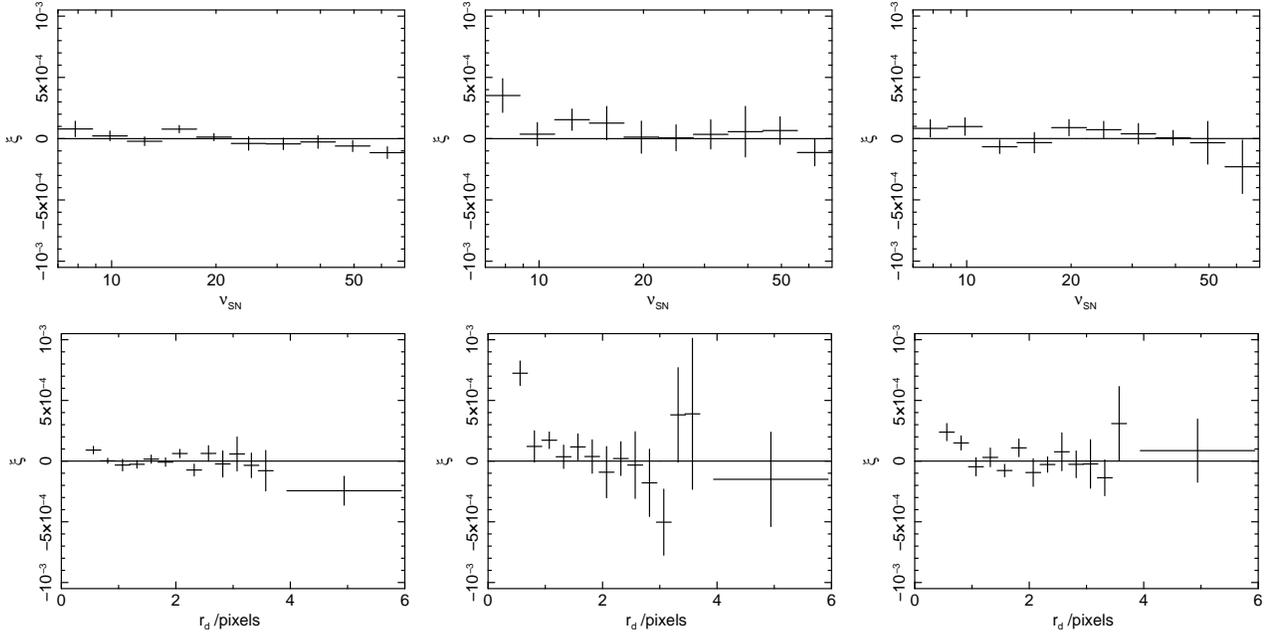

\resizebox{0.3\textwidth}{!}{
\rotatebox{-90}{
\includegraphics{W1m2p1_i_v7p0p3_10.0_24.7.newellip.asc.sn.ps}
}}\hspace*{3mm}
\resizebox{0.3\textwidth}{!}{
\rotatebox{-90}{
\includegraphics{W1p3m0_i_v7p0p3_10.0_24.7.newellip.asc.sn.ps}
}}\hspace*{3mm}
\resizebox{0.3\textwidth}{!}{
\rotatebox{-90}{
\includegraphics{W1p2m3_i_v7p0p3_10.0_24.7.newellip.asc.sn.ps}
}}

\hspace*{1mm}
\resizebox{0.305\textwidth}{!}{
\rotatebox{-90}{
\includegraphics{W1m2p1_i_v7p0p3_10.0_24.7.newellip.asc.g12.ps}
}}\hspace*{2mm}
\resizebox{0.305\textwidth}{!}{
\rotatebox{-90}{
\includegraphics{W1p3m0_i_v7p0p3_10.0_24.7.newellip.asc.g12.ps}
}}\hspace*{2mm}
\resizebox{0.305\textwidth}{!}{
\rotatebox{-90}{
\includegraphics{W1p2m3_i_v7p0p3_10.0_24.7.newellip.asc.g12.ps}
}}
\caption{
Examples of cross-correlation between PSF model ellipticity and galaxy ellipticity in three CFHTLenS fields,
from left to right, W1m2p1, W1p3m0 and W1p2m3. Upper panels show the cross-correlation as a function
of signal-to-noise ratio $\nu_{\rm SN}$, 
lower panels as a function of fitted galaxy major-axis scalelength $r_{\rm d}$. 
Vertical bars indicate 68\,percent confidence intervals.
\label{fig:cross}
}
\end{figure*}

To test for these effects, we first measured, in each individual CFHTLenS field, the
cross-correlation between galaxy ellipticity and the mean ellipticity of the PSF model, measured at the
location of each galaxy, with PSF ellipticity averaged over the
exposures used.  The cross-correlation is shown for
three typical fields, as a function of signal-to-noise ratio $\nu_{\rm SN}$ and fitted galaxy size, in Fig.\,\ref{fig:cross}.
Statistical uncertainties were calculated by randomly dividing the observed samples into ten subsamples, and
measuring the variance in the statistic between the subsamples.  The reported error is then $\sqrt{10}$ smaller
than the subsample rms.  However, on scales of one degree, there is a significant additional term that
arises from the random cross-correlation of the cosmic shear signal with the PSF.  In \citet{heymans12a} this
term is explicitly estimated using shear derived from the dark matter simulations of \citet{harnois-deraps12a}. 
In this paper, this additional term is ignored.  
It manifests itself as an uncertainty, which is highly correlated between the plotted
points, to be added in quadrature to the statistical uncertainties shown in Fig.\,\ref{fig:cross}.

The test presented may not be sensitive to all conceivable forms of PSF modelling error.  For example, the PSF 
model may be too small in some regions, leading to positive PSF-galaxy correlation, but too large in others,
leading to negative PSF-galaxy correlation.  Averaging over the field may lead to no net detection of
an effect.  However, we are not able to analyse the fields in smaller subsamples, as the statistical and
cosmic shear uncertainties become large.

The results from the full-field analyses are generally encouraging, with no obvious dependence of a
cross-correlation signal on either signal-to-noise ratio or fitted galaxy size.  The possible exception 
to this statement is at the lowest fitted galaxy scalelengths, less than 0.7\,pixels, where often a
positive cross-correlation is observed.  This is very likely due to reaching the limits of the sampled
models that were generated as described in Appendix\,\ref{appendixA}.  It should not be surprising that systematic
effects may be seen for galaxies with major axis scalelengths smaller than one pixel in size, and considerably less
than the PSF size.  Fortunately, the number of such galaxies is relatively small, and they all have small
weights (Fig.\,\ref{fig:weights}).

If there are residual errors in the relative astrometry, this would result in systematic shear patterns being induced
with scalelengths of variation determined by the scalelength of the astrometric errors (probably many arcminutes).
The resulting systematic signal would likely not be correlated with any other measurable quantity.  We tested for
this in a subset of CFHTLenS fields by switching off the relative astrometry correction described in 
Section\,\ref{image_combination}, remeasuring those fields and generating shear-shear correlation functions.
Only small differences, much smaller than the statistical uncertainties in the correlation function, were observed,
indicating that for CFHTLenS the absolute astrometric accuracy already is adequate for shear measurement from
individual exposures.

An in-depth analysis of the survey for systematic errors is presented by \citet{heymans12a}.

\section{CFHTLenS Image simulations}\label{sims}
The {\em lens}fit shape measurement algorithm has previously undergone a series of verification tests using simulated sheared images that test a range of different observing conditions, galaxy and PSF types \citep{kitching08a}.  These tests have used both sets of simulations from the Shear TEsting Programme \citep[STEP;][]{STEP1, STEP2}, demonstrating sub-percent level accuracy.
The philosophy of {\em lens}fit, to analyse data from individual exposures rather than a co-added image, requires a suite of specialized image simulations that mimic this main feature of the CFHTLenS analysis.  We describe in this section how these simulations compare to previous image analysis challenges and the subsequent verification and calibration analysis.   

As well as confirming that the algorithms described above do work on multiple exposures, 
the simulations also allow us to measure
the calibration term that is required to correct for noise bias, as discussed in Section\,\ref{shear}, in
Appendix\,\ref{appendixC} and by \citet{refregier12a} and \citet{melchior12a}.  \citet{kacprzak12a} also describe the
use of simulations for calibration of noise bias.

\subsection{GREAT CFHTLenS Image Simulations}

Our philosophy followed the strategy of the GREAT challenges\footnote{www.greatchallenges.org} \citep{bridle10a, GREAT10}, to design the simplest simulation that nonetheless addresses the specific question: how accurate is {\em lens}fit when analysing realistic galaxy distributions, measured with multiple, low-signal-to-noise exposures?   We therefore chose a constant PSF across the field of view, and we postpone discussion of our tests on the robustness of the measurements to a spatially varying PSF in the actual data to the companion paper \citep{heymans12a}.  In the simulations, we did, however, vary the PSF between the multiple exposures simulated for each field, based on the variation between exposures that we see in the CFHTLenS data.

We simulated the full CFHTLenS area with each simulated one square degree field consisting of seven dithered exposures, with a linear dithering pattern that was matched to the average linear dither pattern in the survey, ignoring higher order astrometric corrections.  Galaxies and stars were positioned within the field with the same distribution as in the real data and with matching magnitude and signal-to-noise ratio.    We then created a population of disk-plus-bulge galaxies such that the distributions of galaxy brightness, size, intrinsic ellipticity and bulge fraction matched the distributions described in Section\,\ref{galaxymodels} and Appendix\,\ref{appendixB}.  
As in those models, the disk and bulge ellipticities were set equal.
The unlensed orientation of each galaxy was random and each galaxy was assigned a cosmological shear based on its original position and redshift in the CFHTLenS data, as simulated by our N-body lensing simulation of CFHTLenS \citep{harnois-deraps12a}.  In order to potentially allow reduction of the impact of noise from the intrinsic galaxy ellipticity, we followed \cite{STEP2} by creating two image simulations for each field, with the intrinsic ellipticities rotated by $90^{\circ}$ between the two images, although in practice we found that such shape noise mitigation was itself problematic and was therefore not used, as discussed in Section\,\ref{simanalysis}.  Camera distortions were not included in the simulations but the {\em lens}fit pipeline still applied a correction as it would on real data.

\begin{figure}
   \centering
   \includegraphics[width=3.4in, angle=0]{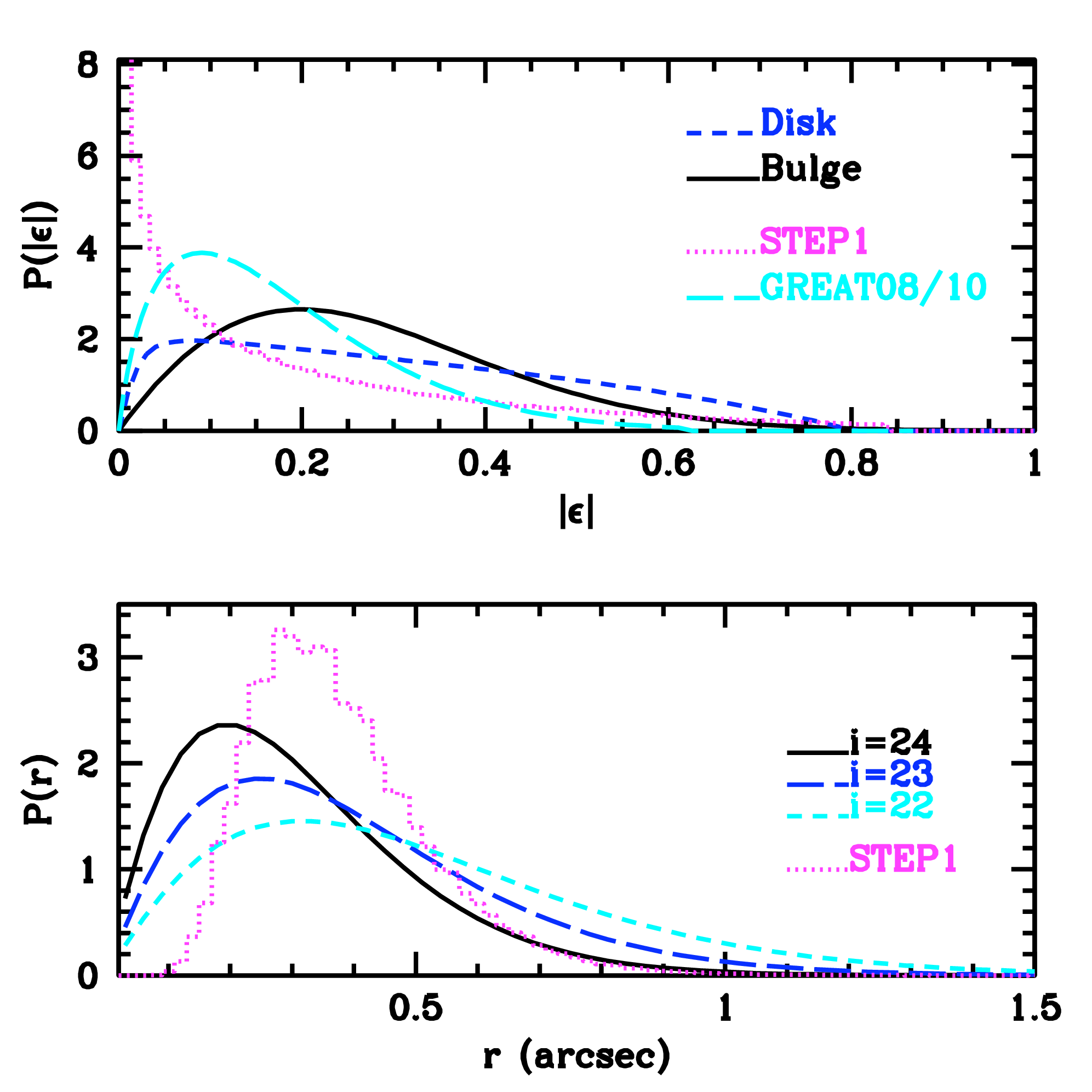} 
   \caption{ 
Upper panel: the intrinsic ellipticity
distribution for disk-dominated galaxies in the STEP (dotted curve), 
GREAT (long-dashed) and CFHTLenS (short dashed) simulations and for
bulge-dominated galaxies in the CFHTLenS simulations (solid curve).
Lower panel: the magnitude-dependent galaxy size distribution,
where $r$ is the major-axis half-light radius for the bulge component
and the major-axis exponential scalelength for the disk component.
The CFTHLenS simulated distributions are shown at magnitudes $i'=22$ 
(short dashed curve), $i'=23$ (long dashed) and $i'=24$ (solid).
The distribution of the first STEP challenge is also shown (dotted curve).
\label{fig:priors}
}
\end{figure}

The model prior distributions differed significantly from the distributions
assumed in previous image challenges, as shown in
Fig.~\ref{fig:priors}.  We simulated the galaxy population as comprising
two types: 90\,percent of galaxies were disk-dominated, the remainder
were bulge-dominated.  No irregular galaxies were included.
The upper panel compares the intrinsic ellipticity distributions in
our custom CFHTLenS simulations to those simulated in the first STEP
challenge and both GREAT challenges.  It can be seen that the distributions
of ellipticity for the disk-dominated population differ significantly.
The bulge-dominated galaxy population
ellipticity distribution was more similar to the ellipticity
distribution assumed in previous challenges, although the fraction
of bulge-dominated galaxies in the STEP simulations was
negligible.  The lower panel of Fig.~\ref{fig:priors} shows the size
distribution for both populations, which varied with galaxy magnitude in our custom
simulations.  This can be compared to the somewhat larger size
distribution for the first STEP challenge, scaled according to the
relative pixel sizes in STEP and CFHTLenS, and the set of three
discrete galaxy-to-PSF size ratios simulated in the GREAT08 challenge,
which for the average CFHTLenS PSF corresponds to galaxy scalelengths
of $r=[0.2,0.3,0.6]$ arcsec.  Comparing these distributions indicates
how challenging our custom simulations were compared to previous image
simulations, owing to the more realistic, broad ellipticity
distributions and high fraction of small galaxies. We therefore would
expect neither {\em lens}fit nor other methods to perform as well on
these custom simulations as in previous challenges.

The image simulations were produced using a modified version of the
GREAT challenges image simulation software package detailed in
Appendix\,A.3 of \citet{GREAT08}.  Aside from the differences in
galaxy populations described above, our simulations also differed from
the GREAT08 and GREAT10 simulations by allowing for overlapping object
isophotes, by having each exposure containing both galaxy and stellar
objects and by simulating dithered data.  We also modified the
definition of galaxy size to be the size of the semi-major axis of a
galaxy's isophotes, instead of the geometric mean of the semi-major and semi-minor axes.
This redefinition is consistent with the convention used for
measurements of data \citep[e.g.][]{simard02a} and avoids 
introducing an incorrect correlation between size and ellipticity 
(see also Appendix\,\ref{appendixA}).
The definition of
signal-to-noise ratio was chosen such that it was defined within a
circular aperture of radius of $0.935''$ (5 pixels) for all galaxies.
Using a matched aperture flux, as measured by SExtractor on the data,
then allowed us to directly replicate the signal-to-noise
distributions within each CFHTLenS exposure.

\begin{figure}%  figure placement: here, top, bottom, or page
   \centering
   \includegraphics[width=3.4in, angle=0]{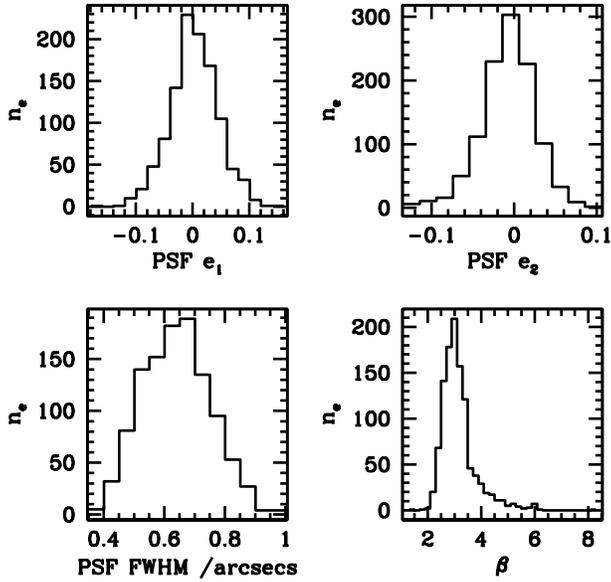} 
   \caption{The distribution of PSF parameters measured from the CFHTLenS data and replicated in the CFHTLenS simulations.  The upper panels show the PSF ellipticity.  The lower panels shows the PSF FWHM in arcsec (left) and the shape parameter $\beta_{\rm Moffat}$ (right).}
   \label{fig:psfparams}
\end{figure}

The choice of simulated PSF model was motivated by the data.  To each of the 36 MegaCam CCDs in each CFHTLS exposure, we fitted an elliptical-isophote \citet{moffat69a} profile to the images of stellar objects.  A circular light profile at distance $r$ from the centre of a star was given by
\be
I(r) = I_0 \left [ 1 + \left(\frac{r}{r_s}\right)^2 \right] ^{-\beta_{\rm Moffat}}
\ee
where $I_0$ is the peak flux, and the scale radius $r_s$ and the shape parameter $\beta_{\rm Moffat}$ were related to the FWHM of the PSF as
\be
{\rm FWHM} = 2 r_s \, \sqrt{2^{1/ \beta_{\rm Moffat}} - 1}.
\ee
We created a series of circular model PSFs on a pixel grid for a range of scale radii $r_s$ and shape parameters $\beta_{\rm Moffat}$ and applied a linear transformation of the co-ordinates to shear the model PSF to have elliptical isophotes.  We then  calculated the best-fit PSF ellipticity and the Moffat profile parameters $\beta_{\rm Moffat}$ and $\alpha$ for each CCD chip.   For each simulated exposure we adopted a PSF model for the full field of view by selecting a CCD PSF at random.  This was to ensure that we fully represented the range of PSF models at all locations in the images, rather than taking the often more circular average over the full field of view.  We then grouped the PSF model exposures into fields following the grouping of exposures in the real data.  The distribution of PSF parameters measured from the CFHTLenS data, and replicated in the simulations, is shown in Fig.~\ref{fig:psfparams}.

\subsection{SkyMaker CFHTLenS image simulations}\label{skymaker}
When constructing calibration corrections, it is of crucial importance to verify the base image simulations used to ensure the calibration is not dependent on the simulation software used.  To this end we created a second suite of 140 MegaCam image simulations generated using SkyMaker\footnote{www.astromatic.net} \citep{bertin09a}.   
The {\bf SkyMaker} PSFs were not Moffat profiles, but instead 
had radial surface brightness profiles that took into account the telescope optics, 
atmospheric seeing and scattering into wings of the PSF, 
although those adopted here had circular symmetry.  Unlike the GREAT simulations,
only a single exposure was simulated, for reasons of computational time, and the 
signal-to-noise distribution of galaxies on those single exposures 
was matched to that of the CFHTLenS galaxies on the coadded data. 
The signal-to-noise ratios were generally smaller than in the SkyMaker simulations tested in 
the first STEP challenge \citep{STEP1}.   Furthermore, the size and ellipticity distributions were chosen to match those in the custom CFHTLenS simulations described above, unlike the SkyMaker simulations tested in \citet{STEP1}, which follow the dotted curves in Fig.~\ref{fig:priors}.  The galaxy and star positions were chosen at random and circular PSFs with a FWHM of $0.7''$ were used.  A cosmic shear signal from a redshift plane at $z=1$ was assumed in the simulation.

\subsection{Model bias}\label{modelbias}
The primary aim of the simulations was to measure the bias caused by
noise (Appendix\,C), and for this aim the simulated galaxies should
match the real population as closely as possible.
The strategy of matching the
simulations to the models and priors used in the {\em lens}fit
measurements is reasonable, as these distributions have been derived
from observations and are therefore a good representation of the
galaxy population. The tests presented here do not test for the effect
of deviations of the true population from the model assumptions,
but they do test the efficacy of the method taking into account
the implementation of the statistical algorithm in
Section\,\ref{likel}, including the effects of convolution with
appropriate PSFs, the effects of finite signal-to-noise and noise
bias, the combination of multiple dithered images and the effect of
masks applied to the data. 

The difficulty of attempting also to test
for the effect of deviations of the true population from the assumed population is
that we have little prospect of obtaining an independent assessment of
how different the real Universe might be from our assumptions, even at
the level of knowing the distributions of surface brightness of
individual galaxies in the local universe \citep{graham08a}.
Notwithstanding that difficulty, this question has been considered by other authors.
\citet{voigt10a} considered the multiplicative biases that might arise from 
fitting multiple gaussian components to exponential or de~Vaucouleurs surface brightness distributions,
convolved with a PSF.  It was found that fitting a single gaussian to an exponential distribution
caused a multiplicative bias of 3\,percent and an additive bias of $\sim0.001$, but that these biases 
fell to 0.3\,percent and $<10^{-4}$ when two gaussian components were adopted.  Larger biases were
found when fitting gaussians to a pure de~Vaucouleurs profile ($3-6$\,percent and 0.001 respectively, 
when fitting two gaussians),
but this is not surprising given the significant difference in surface brightness profile between
gaussian and de~Vaucouleurs distributions.  We note that pure de~Vaucouleurs-profile galaxies are expected to be 
a small fraction of the true population.  Multiplicative biases of $2-5$\,percent were found when fitting two
gaussians to composite galaxy profiles comprising exponential plus de~Vaucouleurs components: 
in this case we expect our fitting of exponential plus de~Vaucouleurs models to work well and therefore
to lead to significantly smaller biases.

\citet{voigt10a} argue that these biases arise because a gaussian profile is 
too flat to reproduce well the central regions of realistic galaxy profiles.  
In this paper, we fit two components, exponential plus de~Vaucouleurs,
to the real population of galaxies, and these should be far better representations of the true
surface brightness distributions than multiple gaussians, with sufficient flexibility to model well a wide range of
Sers\'ic indices.  We therefore expect model bias to be
significantly smaller than found by \citet{voigt10a} when fitting gaussians, and to be
subdominant compared with noise bias, as evaluated in the following sections, in the CFHTLenS analysis.

Future surveys covering larger sky areas will have more stringent requirements on the acceptable level
of systematics than CFHTLenS, and it is likely that the question of model bias will need to be
revisited for those surveys.

\subsection{Analysis and Calibration}\label{simanalysis}
\begin{figure}%  figure placement: here, top, bottom, or page
   \centering
   \resizebox{0.48\textwidth}{!}{
   \includegraphics{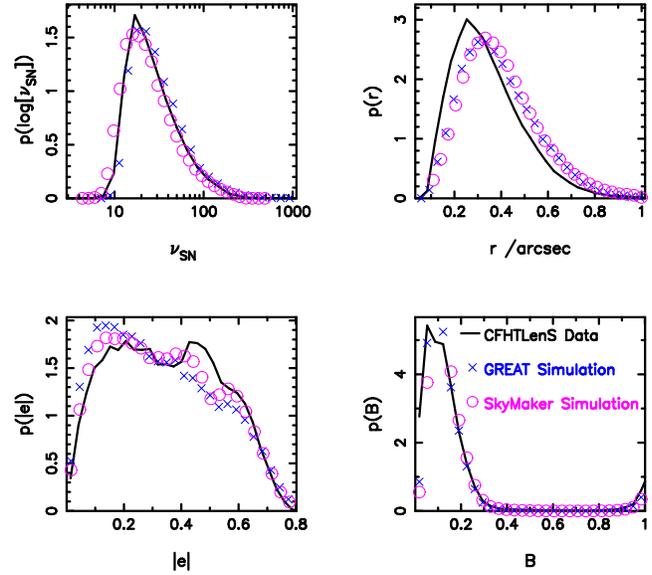} 
   }
   \caption{Comparison of the weighted distributions of measured signal-to-noise ratio (upper left), galaxy size (upper right),  ellipticity distributions (lower left) and bulge fraction (lower right) from the custom GREAT image simulations (crosses), custom SkyMaker image simulations (circles) and the CFHTLenS data (solid).}
   \label{fig:comp_w_data}
\end{figure}

We processed both sets of image simulations with {\em lens}fit using
the known galaxy and star positions as an input catalogue.
Fig.\,\ref{fig:comp_w_data} shows the measured galaxy size,
signal-to-noise ratio, bulge fraction and ellipticity distributions from
both sets of image simulations and the CFHTLenS data.  In all cases
the distributions incorporate the weights assigned by {\em lens}fit
(equation\,\ref{eqn:weight}).  The simulated signal-to-noise
distribution shown in the upper left panel was constructed to match
the CFHTLenS depth.  We found reasonable agreement between the ellipticity
distribution measured in the data and simulations, although in the weighted distributions of CFHTLenS there is
an excess of galaxies at $| \bmath{e} | \simeq 0.45$. We do not expect perfect agreement between the measured
distributions and priors, because only the summed posterior distribution is expected to match the prior distribution
(Paper I), whereas Fig.\,\ref{fig:comp_w_data} shows the values obtained from the expectation values of
the probability distribution (after marginalising over the other parameters).  The excess is only present
in the weighted distribution, and not in the unweighted distribution, which implies that, on average, galaxies
with non-zero, middling ellipticities have more sharply defined likelihood surfaces than their peers at
lower or higher ellipticity.  This effect is not fully understood, but it is probably linked to the issue of noise
bias that is discussed in Appendix\,\ref{appendixC}.
In comparison to the data, in the simulations there are also lower fractions of
small galaxies for which an ellipticity was determined.  As these small
galaxies do exist in the simulation (see Fig.\,\ref{fig:priors}) this
might indicate a mismatch between the PSF in the data and simulations, such
that small galaxies were harder to measure in the simulations than in the real data.
Either pixelisation and noise from the data led us to overestimate the
PSF size which was then used in the simulation, or the Moffat profile
used in the GREAT simulations and the SkyMaker PSF 
were poor models of the MegaCam PSF.  There are, however, sufficient numbers of
small galaxies to calculate a calibration as described below, and as
these small galaxies were assigned the lowest weights by application
of equation\,\ref{eqn:weight}, it should not be problematic that their
calibration was based on relatively fewer galaxies than for other
sizes in the data.

We follow \citet{STEP1} in modeling our shape measurement as a linear
combination of a multiplicative error $m$ and an additive error $c$
such that \be \bmath{e}^{\rm obs} = (1 + {\rm m}) \bmath{e}^{\rm true}
+ {\bf c} + \Delta \bmath{e} \ee where $\bmath{e}^{\rm obs} $ is the
observed complex ellipticity as measured by {\em lens}fit, $\Delta
\bmath{e}$ is the noise on the measure and $\bmath{e}^{\rm true}$ is
the sheared intrinsic ellipticity defined as in
equation\,\ref{eqn:shear} and for simplicity we treat $m$ as an
orientation-independent scalar.  At
this point we are faced with several alternatives to determine $m$ and
$\bmath{c}$.  Comparing $e^{\rm obs}$ directly with $e^{\rm true}$
would lead to a strong noise bias in the measurement of $m$ and
$\bmath{c}$, as the observed ellipticity is bounded with
$|\bmath{e}^{\rm obs}| < 1$ such that $\langle \Delta \bmath{e}
\rangle$ is non-zero for a galaxy with non-zero ellipticity.  Comparing
$\bmath{e}^{\rm obs}$ directly with shear $\bmath{\gamma}$ would also
lead to an intrinsic ellipticity noise bias, as for a finite number of
galaxies $\langle \bmath{e}^{\rm true} \rangle$ may be non-zero.  This
is particularly relevant when galaxies are binned in order to look for
size- and signal-to-noise-dependent calibrations, increasing the
random shot noise in each bin.  \citet{STEP2} developed a method to
circumvent this issue by including rotated pairs of galaxies in the
simulations such that $\langle \bmath{e}^{\rm int} \rangle = 0$ by
construction, where $\bmath{e}^{\rm int}$ is the intrinsic ellipticity
prior to application of any lensing shear.  We generated such galaxy
pairs in the custom GREAT simulations, but found that, when galaxies
within a pair were assigned different weights, which at the low
signal-to-noise regime was typical, 
$\langle w \bmath{e}^{\rm int}\rangle$ was significantly non-zero.
Thus we did not use this method of shape-noise
cancellation. Instead, our preferred analysis method was to first bin
galaxies by size $r$ and
signal-to-noise ratio $\nu_{\rm SN}$, such that there were equal
numbers of objects in each bin, and then to subdivide those bins by the input cosmic shear
of each galaxy.
Each galaxy was entered into two bins of shear, treating each component as being an independent
measure of shear for the purposes of deriving the calibration correction.
Then, a calibration was derived by minimizing
$\chi^2$ given by 
\be \chi^2 = \sum_i \frac{1}{\sigma_\gamma^2} \left(\bar{\gamma}_{i}^{\rm obs} - \left[ 1 + {\rm m}(\nu_{\rm SN},r)
  \right] \bar{\gamma}_{i}^{\rm true} - {\rm c}(\nu_{\rm SN},r)\right)^2 \, , \ee 
where for each bin $i$ 
\begin{equation}
\bar{\gamma}_{i}^{\rm obs} = \frac{\sum_{j} w_j ({e}_{\alpha j}^{\rm obs} - {e}_{\alpha j}^{\rm int})}{\sum_{j} w_j}
\label{eqn:gobs}
\end{equation}
and where for galaxy $j$ in bin $i$, ${e}_{\alpha j}$ is either the ${e}_1$ or ${e}_2$ component of the ellipticity, with $\alpha=1,2$ being set by the component of the input cosmic shear for that galaxy in that bin.

The error $\sigma_\gamma$ on $\bar{\gamma}_{i}^{\rm obs}$ was
estimated through a bootstrap error analysis within each bin $i$.
$\sigma_\gamma$ was found to be uncorrelated between bins. 
The summation over both ellipticity components
in equation~\ref{eqn:gobs} was motivated by an
initial component-dependent minimization, which showed negligible
differences in the calibration measured for each ellipticity
component.

\begin{figure}%  figure placement: here, top, bottom, or page
   \centering
   \resizebox{0.48\textwidth}{!}{
     \rotatebox{-90}{
   \includegraphics{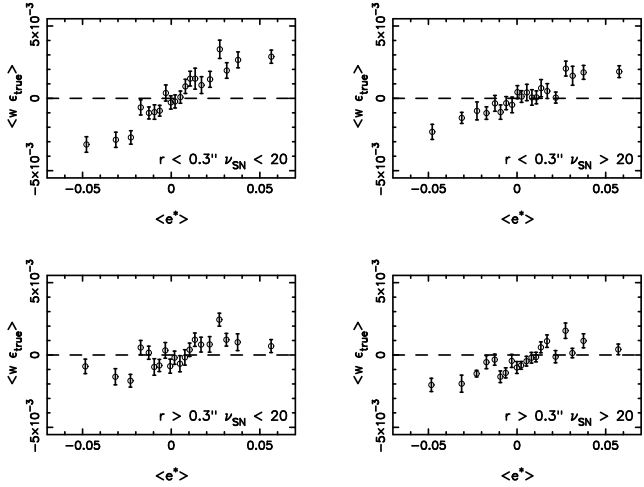}}}
   \caption{The weighted average true intrinsic ellipticity $\langle w e^{\rm int} \rangle$ as a function of the PSF ellipticity $e^*$ for four sets of GREAT-simulated galaxies binned by galaxy size and signal-to-noise ratio.  Small galaxies with major-axis scalelengths $r<0.3''$ are shown in the upper panels; large galaxies with $r>0.3''$ are shown in the lower panels.  The left-hand panels show the faint sample with $\nu_{\rm SN}<20$; 
the brighter sample with $\nu_{\rm SN}>20$ are shown on the right-hand panels.  Each bin contains roughly the same number of galaxies.}
   \label{fig:weighteint}
\end{figure}

This analysis method ensured that we were unbiased to measurement or intrinsic ellipticity noise, but it does disguise biases introduced by the weights.  In Fig.\,\ref{fig:weighteint} we use the GREAT CFHTLenS simulations to show the weighted average true intrinsic ellipticity $\langle w e^{\rm int} \rangle$ as a function of the PSF ellipticity $e^*$ for four sets of galaxies binned by galaxy size and signal-to-noise ratio.  For each galaxy sample we find a preferential weighting of galaxies oriented in the same direction as the PSF.  This weight bias is independent of which ellipticity component we choose and affects bright large galaxies to a similar extent to small, faint galaxies.  It is an undesirable outcome of the weighting scheme, which correctly identifies that it is easier to measure the shapes of galaxies oriented with respect to the PSF in contrast to galaxies oriented perpendicular to the PSF, and may be viewed as a form of the orientation bias discussed by \citet[][Section\,\ref{weights}]{bernstein02a}.  We investigated isotropisation schemes to remove the PSF-related weight bias, by assigning galaxies an averaged weight based on its size relative to the PSF size and the signal-to-noise ratio of the galaxy.  We found however that the isotropisation reduced the effectiveness of the weights, which also decrease the noise bias in the shear measurement (Appendix\,\ref{appendixC}), and resulted in an even stronger weight bias in the opposite sense.  We therefore concluded that weight bias, which impacts shape measurement at the $10^{-3}$ level, is an undesirable feature of applying inverse-variance weighting, but is a small effect for the study of CFHTLenS.  In \citet{heymans12a} we describe a method to identify and remove fields which show any significant correlation with the PSF, which can arise both from this weight bias and from the general effect of noise bias.

\begin{figure}%  figure placement: here, top, bottom, or page
   \centering
   \resizebox{0.48\textwidth}{!}{
   \includegraphics{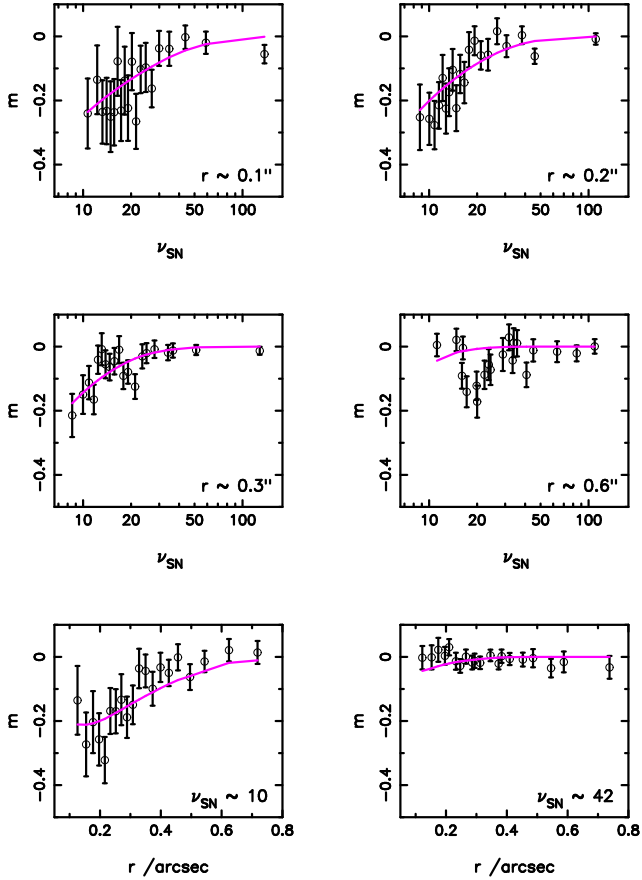} }
   \caption{Example results of the measured multiplicative error ${\rm m}(\nu_{\rm SN},r)$ 
in the GREAT simulations as a function of signal-to-noise ratio for four size bins selected to be the smallest scale length ($0.1''$, upper left), and the three discrete galaxy sizes equivalent to those simulated in GREAT08 (0.2, 0.3, 0.6 arcsec).  The solid curve is the best fitting functional form to the full data set which comprises 20 bins in both signal-to-noise ratio and size.
The lower two panels show ${\rm m}(\nu_{\rm SN},r)$ as a function of $r$ for two bins of $\nu_{\rm SN}$.
}
   \label{fig:mcSN}
\end{figure}

\begin{figure}%  figure placement: here, top, bottom, or page
   \centering
   \resizebox{0.23\textwidth}{!}{
   \includegraphics{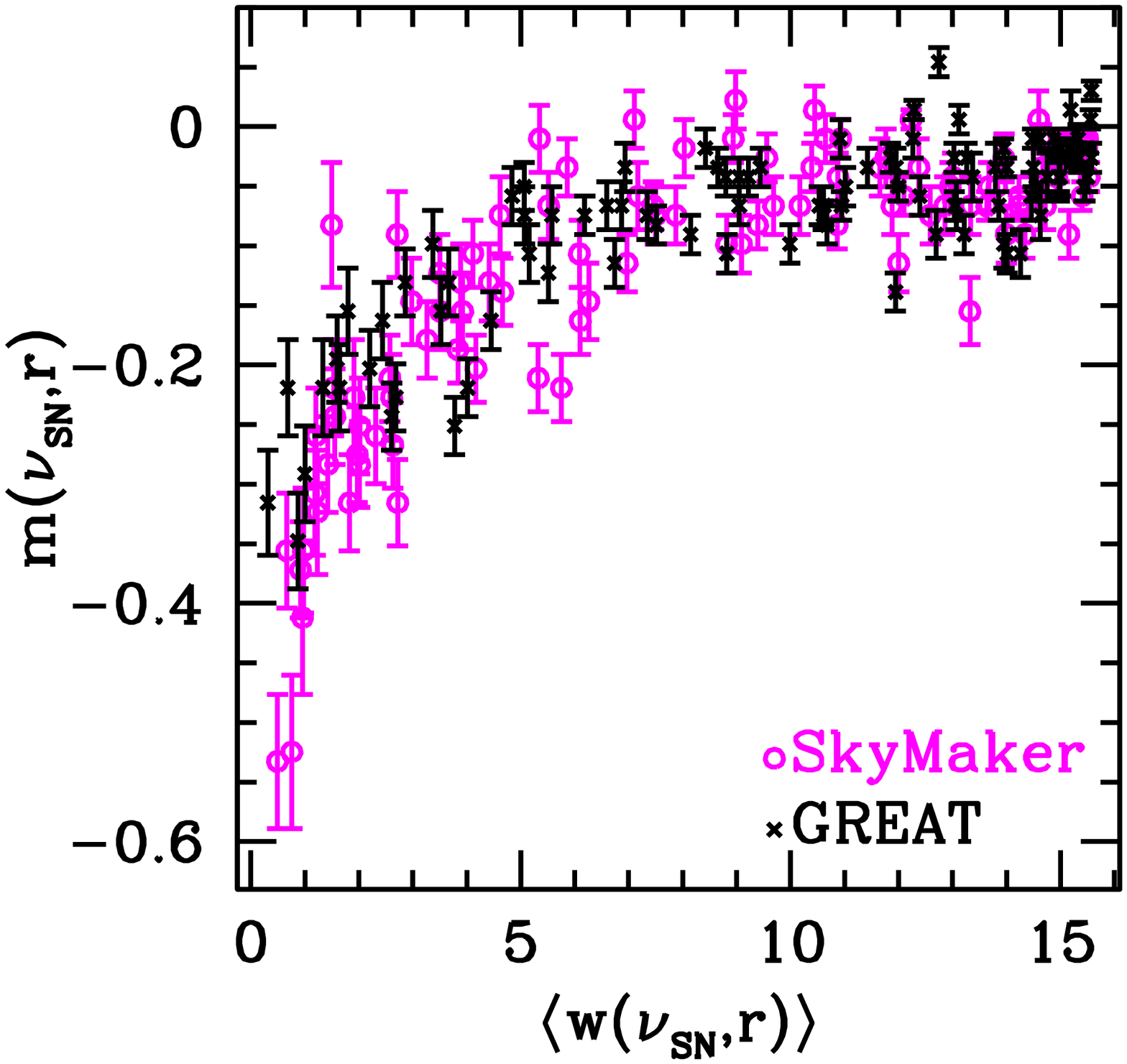}}
   \resizebox{0.23\textwidth}{!}{
   \includegraphics{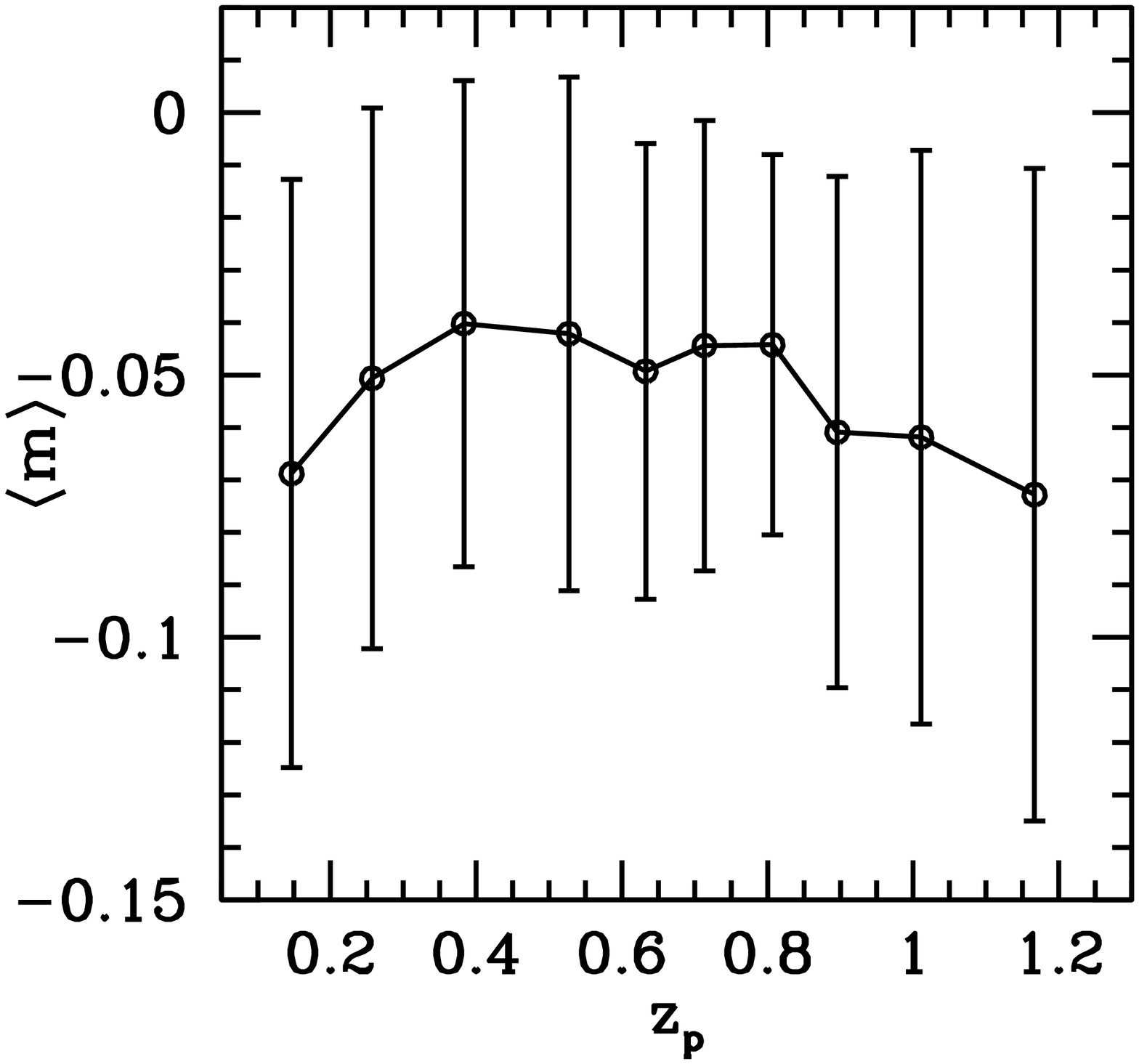}}
   \caption{
(left) The multiplicative calibration $m$ for each signal-to-noise ratio and size bin 
as a function of the average {\em lens}fit weight $\langle w \rangle$ of the simulated galaxies in the bin.
Results are shown from the GREAT (crosses) and the SkyMaker (open circles) simulations.
(right) The mean applied multiplicative calibration $\langle m \rangle$ for CFHTLenS galaxies binned 
by photometric redshift $z_p$.  Vertical bars indicate the rms of the distribution of $m$ values
within each bin.
   \label{fig:mvsweight}
}
\end{figure}

Fig.\,\ref{fig:mcSN} shows some example results of the measured multiplicative error ${\rm m}(\nu_{\rm SN},r)$ 
in the GREAT simulations as a function of $\nu_{\rm SN}$ for four size bins selected to be the smallest, and the three discrete galaxy sizes simulated in GREAT08.  The solid curve is a fit to the full data set, which comprises 20 bins in each of $\nu_{\rm SN}$ and size, given by
\be
m(\nu_{\rm SN},r) = \frac{\beta}{\log(\nu_{\rm SN})} \exp\left(-\alpha r \, \nu_{\rm SN} \, \right)
\label{eqn:alphabeta}
\ee
with $\alpha = 0.057$ and $\beta=-0.37$.  We find that for all galaxies with $r>0.5''$, or $\nu_{\rm SN}>35$, $m$ is consistent with zero on average, 
but that the calibration correction increases for smaller and lower signal-to-noise galaxies,
as expected for noise bias (Appendix\,\ref{appendixC}).   The {\em lens}fit weighting scheme already downweights these noisy galaxies, such that even though the calibration correction can be significant, these galaxies contain very little weight in our scientific analyses.  This can be seen in the left panel of Fig.\,\ref{fig:mvsweight}, which shows the multiplicative calibration for each bin of signal-to-noise ratio and size as a function of the average weight of the galaxies in the bin, for both the GREAT and
SkyMaker simulations.  It can be seen that the net effect is very similar in the two sets of simulations, although the
SkyMaker simulations display a slightly larger bias than the GREAT simulations for galaxies with low signal-to-noise
ratio and thus lower {\em lens}fit weight.  As the GREAT simulations have anisotropic PSFs that
closely match the CFHTLenS PSFs, and as they incorporate the effect of having seven multiple exposures,
whereas the SkyMaker simulations are single exposures in each field, we use the results from the GREAT simulations to
calibrate the CFHTLenS shear measurements.  The differences between the results from the simulations only become
apparent for galaxies which are significantly downweighted in CFHTLenS, and overall it is encouraging to note the 
close match between the two sets of simulations.  The right panel of Fig.\,\ref{fig:mvsweight} shows how the mean applied
calibration correction $\langle m \rangle$ varies with the photometric redshift of the galaxies in CFHTLenS, showing
primarily the larger correction that needs to be applied at redshifts $z_p \ga 0.8$ owing to the decreasing 
signal-to-noise ratios and sizes of galaxies at higher redshifts.  We note also the larger correction deduced for galaxies
at low redshifts, $z_p \la 0.3$, which probably arises from contamination of the low-redshift population by 
galaxies with higher spectroscopic redshifts \citep[see][for further discussion]{heymans12a}.

We find no evidence for a significant additive error $\bmath{c}$ term in the {\em lens}fit analysis of the simulations, 
indicating that the multiplicative error arising from noise bias is the dominant term to be corrected.
 
\subsection{Simulated Cosmic Shear Tomography}
\begin{figure}%  figure placement: here, top, bottom, or page
   \centering
   \resizebox{0.48\textwidth}{!}{
   \includegraphics{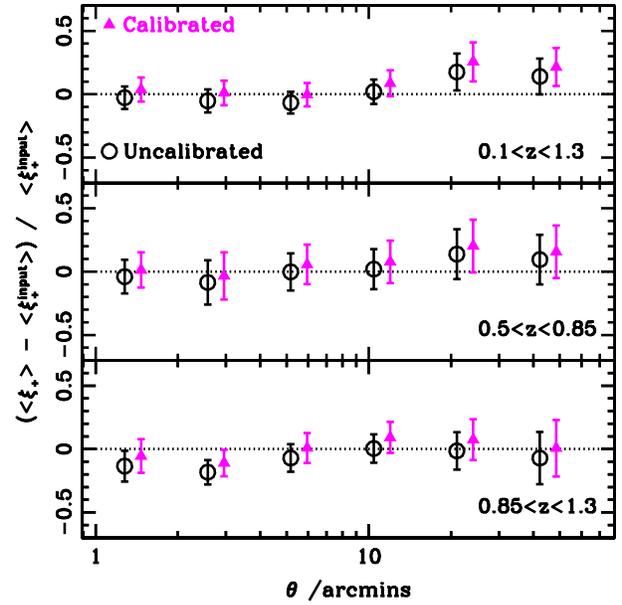} }
   \caption{Comparison of the measured to input two-point correlation function for the raw uncalibrated data (circles) and the calibrated data (triangles) from the GREAT CFHTLenS simulations.   The error bars are representative of the signal-to-noise ratio expected for a CFHTLenS-like survey.  To investigate the impact of the calibration as a function of redshift we show the signal for three tomographic configurations; $0.1<z<1.3$ (upper panel), $0.5<z<0.85$ (middle panel) and $0.85<z<1.3$ (lower panel). } 
   \label{fig:xiratio}
\end{figure}

To test the application of such a calibration, we estimated the calibration-corrected 
two-point shear correlation function
(other papers will describe the correction to estimators such as the three-point function and others).
We first calculated the uncalibrated two-point function,
\be
\xi_{\pm}(\theta) = \frac{\sum w_i w_j \left[ \epsilon_t (\bm{x_i}) \epsilon_t (\bm{x_j}) \, \pm \, \epsilon_r (\bm{x_i}) \epsilon_r (\bm{x_j})
\right]}{
\sum w_i w_j } \, .
\label{eqn:xipm_est}
\ee
where the weighted sum is taken over galaxy pairs with $|\bm{x_i - x_j}|=\theta$.  The two-point correction term was then estimated as
\be
1 + K(\theta) = \frac{\sum w_i w_j [1+m(\nu_{{\rm SN}, i},r_i)][1+m(\nu_{{\rm SN}, j},r_j)]}{
\sum w_i w_j } \, .
\ee
for the same galaxy pairs, thus taking into account potential astrophysical correlations between galaxy size, brightness and clustering scales.  The calibration corrected  two-point function was then given by
\be
\xi_{\pm}^{\rm cal} (\theta) = \frac{\xi_{\pm}(\theta)}{1 + K(\theta)}  \, .
\ee
The alternative of applying the calibration correction directly to individual galaxies could result in instability in the rare events where $1+m(\nu_{{\rm SN},i},r_i) \rightarrow 0$.  In addition, using this method automatically removes any unwanted correlation between the calibration correction and intrinsic ellipticity which could arise from the measured galaxy signal-to-noise ratio  or size being weakly correlated with shape, which is a possible outcome of any shape measurement method.  We show a comparison of the measured and input two-point correlation functions in Fig.\,\ref{fig:xiratio}, for the raw uncalibrated data and the calibrated data.   The error bars are representative of the signal-to-noise ratio expected for a CFHTLenS-like survey and the data points are strongly correlated.  To investigate the impact of the calibration as a function of redshift we show the signal for three tomographic configurations; a full sample with $0.1<z<1.3$ similar to the sample used in \citet{kilbinger12a} and two tomographic bins with $0.5<z<0.85$ and $0.85<z<1.3$, similar to the two tomographic bins used in Benjamin et al. (in preparation).  Note that a very low redshift bin with $0.1<z<0.5$ has statistical error bars that encompass the full figure.  Fig.\,\ref{fig:xiratio} shows that for a CFHTLenS-like survey the correction is well within the statistical errors.  We do see that the calibration correction is more significant for the higher redshift sample, as expected given that the average size and signal-to-noise ratio decreases with redshift.  

\begin{figure}%  figure placement: here, top, bottom, or page
   \centering
   \resizebox{0.35\textwidth}{!}{
   \rotatebox{-90}{
   \includegraphics{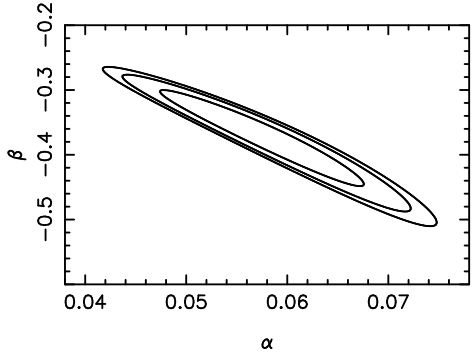} }}
   \caption{Confidence regions of the multiplicative calibration correction parameters in equation~\ref{eqn:alphabeta} used to infer systematic error contribution to the calibration corrected  two-point function $\xi_{\pm}^{\rm cal} (\theta)$.
   \label{fig:alphabeta}
}
\end{figure}

We determined the systematic error contribution to the calibration-corrected two-point function $\xi_{\pm}^{\rm cal} (\theta)$ as follows.  First, we took the likelihood distribution of the multiplicative calibration correction parameters in equation~\ref{eqn:alphabeta}, as constrained by a fit of the calibration correction to the simulated data.  This likelihood distribution is shown in Fig.\,\ref{fig:alphabeta} with the 68\%, 90\% and 95\% confidence intervals.
We randomly sampled this distribution
such that we had a sample of $N=100$ $(\alpha, \beta)$ values that fully represented the probability of the fit (i.e 68\% of the pairs lay within the $1\sigma$ contour).  We then calculated the two-point correction term $K_N(\theta)$ for each of the $N$ resampled values and determined the variance of the calibration-corrected two-point function $\xi_{\pm}^{\rm cal} (\theta)$ between the $N$ differing corrections.  This method allows us to isolate a systematic error covariance matrix for the two-point shear correlation function, allowing marginalisation over the uncertainty on the calibration correction.  
\citet{kilbinger12a} show that this has a negligible effect on cosmological
parameter estimates from the two-point shear correlation function.
The concept can be easily modified to cover any number of the weak lensing statistics that will be used to analyse CFHTLenS.

\section{Conclusions and further developments}\label{future}
We have presented the method of shear measurement developed for CFHTLenS, based on the earlier {\em lens}fit
algorithm of \citet{miller07a} and \citet{kitching08a}.  The method fits PSF-convolved galaxy models
to multiple galaxy images to estimate galaxy lensing shear.
Significant enhancements have been made to the basic algorithm
in order to create a method that can be applied to the survey, as follows.
\begin{enumerate}
\item
Creation of pixel models of the PSF on individual exposures, allowing for spatial and temporal variation of the PSF
and for discontinuities between detectors in the imaging camera.
\item 
Use of two-component models, disk plus bulge, with a prior distribution for bulge fraction estimated from published
work.
\item 
Measurement of likelihood from multiple dithered exposures, optimally combining information when the PSF differs
between exposures, and automatically allowing any one galaxy to have differing numbers of useful exposures, where
that number may vary because of the effect of detector and field boundaries coupled with the dither pattern.  
The multiple exposures are neither interpolated nor co-added, avoiding systematic effects that are otherwise
introduced by those processes.
\item
Measurement of camera distortion from astrometric measurements, and correction of fitted models to optimally remove
the effects of distortion from the shear measurement, rather than correction of the data using interpolation
(i.e. interpolation of data is avoided).
\item
Use of realistic priors on galaxy size, ellipticity and bulge fraction, based on published HST and other data.
\item 
Improved algorithms for bayesian marginalisation over galaxy position, size, flux and bulge fraction.
\item 
Modification of the shear estimator from a posterior probability estimator to a likelihood estimator, to avoid
anisotropy and bias caused by implementation of the sensitivity correction described by \citet{miller07a}, 
but at the cost of needing to calibrate the effect of noise bias on the estimator.
\end{enumerate}

We regard the issue of making measurements from non-interpolated individual exposures as being particularly
important.  If images are interpolated and co-added, the resulting non-stationary convolution kernel that
has been introduced can have a severely detrimental effect on lensing shear measurement.  
In principle, the varying convolving kernel could be evaluated at each galaxy location and a correction applied 
to the PSF, but working from individual exposures avoids that complexity.
Furthermore, the
resulting co-added PSF is likely to have even more complexity than the PSF on individual exposures, and 
in general would be expected to lead to co-added PSFs with `twisted' isophotes (ellipticity and centroid
varying as a function of distance from the peak of the PSF), which
would cause problems for many published methods of shear measurement.  As multiple exposures are usually dithered,
and as the PSF may be discontinuous between detectors (as in MegaCam) and as there are gaps between detectors,
any co-added PSF inevitably has discontinuous variations across the blurred regions between
detectors in the dithered images. The 'gap' regions are not generally sufficiently large to allow independent 
measurement of the PSF from stars in those regions, so either such regions would need to be excised, or the
PSF would need to be measured on individual exposures and the local PSF reconstructed appropriately. 
Excision of such problematic regions would lead to significant reduction in
area for surveys with a large dither pattern. 
These problems are all avoided with the joint-fitting approach described here. 

We have tested the method using two, independent, extensive simulations of CFHTLenS based on the GREAT and SkyMaker
image simulation codes, but with galaxy parameters more closely matched to those found in the faint galaxy
populations than have been assumed in previous lensing challenges.  The faint galaxy population has galaxies
of smaller size and a broader range of ellipticity than has been assumed in the STEP and GREAT challenges, which
leads to a degradation in performance compared with what would have been naively expected from those 
simulations.  When creating image simulations to test shear measurement methods, it is important to include real
observational problems such as those described in this paper, and also to ensure that the simulated
galaxy population mimics closely the observed distribution.  With the current generation of
surveys, we have only been able to match the coarse features of the population, namely size and ellipticity,
and even there uncertainties remain over issues such as the redshift distribution of galaxy properties.
Future, large-area surveys which aim to achieve significantly lower systematic uncertainties than in
CFHTLenS should be concerned about our current lack of knowledge in this area.

Finally, the issue of noise bias has been discussed by a number of recent authors, and with current estimators
we do need to calibrate the noise bias from realistic simulations.  One hope for the future is that a
properly-formulated bayesian estimator may be able to rigorously correct for noise bias without the need
for calibration simulations.

\vspace*{5mm}
\noindent
{\large \bf ACKNOWLEDGMENTS}\\ 
This work is based on observations obtained with MegaPrime/MegaCam, a joint project of CFHT and CEA/DAPNIA, at the Canada-France-Hawaii Telescope (CFHT) which is operated by the National Research Council (NRC) of Canada, the Institut National des Sciences de l'Univers of the Centre National de la Recherche Scientifique (CNRS) of France, and the University of Hawaii. This research used the facilities of the Canadian Astronomy Data Centre operated by the National Research Council of Canada with the support of the Canadian Space Agency.  We thank the CFHT staff for successfully conducting the CFHTLS observations and in particular Jean-Charles Cuillandre and Eugene Magnier for the continuous improvement of the instrument calibration and the Elixir detrended data that we used. We also thank TERAPIX for the individual exposures quality assessment and validation during the CFHTLS data acquisition period, and Emmanuel Bertin for developing some of the software used in this study. CFHTLenS data processing was made possible thanks to significant computing support from the NSERC Research Tools and Instruments grant program, and to HPC specialist Ovidiu Toader. 

CH acknowledges support from the European Research Council under the EC FP7 grant number 240185.
TDK acknowledges support from a Royal Society University Research Fellowship.
LVW acknowledges support from the Natural Sciences and Engineering Research Council of Canada (NSERC) and the Canadian Institute for Advanced Research (CIfAR, Cosmology and Gravity program).
TE is supported by the Deutsche Forschungsgemeinschaft through project ER 327/3-1 and the Transregional Collaborative Research Centre TR 33 - "The Dark Universe".
H. Hildebrandt is supported by the Marie Curie IOF 252760 and by a CITA National Fellowship.
H. Hoekstra acknowledges support from  Marie Curie IRG grant 230924, the Netherlands Organisation for Scientific Research (NWO) grant number 639.042.814 and from the European Research Council under the EC FP7 grant number 279396.
YM acknowledges support from CNRS/INSU (Institut National des Sciences de l'Univers) and the Programme National Galaxies et Cosmologie (PNCG).
JPD was supported by NSF grant AST~0807304.
LF acknowledges support from NSFC grants 11103012 \& 10878003, Innovation Program 12ZZ134 and Chen Guang project 10CG46 of SMEC, STCSM grant 11290706600 and Pujiang Program 12PJ1406700. 
MJH acknowledges support from the Natural Sciences and Engineering Research Council of Canada (NSERC).
TS acknowledges support from NSF through grant AST-0444059-001, SAO through grant GO0-11147A, and NWO.
MV acknowledges support from the Netherlands Organization for Scientific Research (NWO) and from the Beecroft Institute for Particle Astrophysics and Cosmology.

The N-body simulations used in this analysis were performed on the TCS supercomputer at the SciNet HPC Consortium. SciNet is funded by: the Canada Foundation for Innovation under the auspices of Compute Canada; the Government of Ontario; Ontario Research Fund - Research Excellence; and the University of Toronto. 
We thank Sree Balan, Sarah Bridle and 
Emmanuel Bertin for the simulation code used in this analysis.  

{\small Author Contributions: All authors contributed to the development and writing of this paper.    The authorship list reflects the lead authors of this paper (LM, CH, TDK, LVW) followed by two alphabetical groups.  The first alphabetical group includes key contributors to the science analysis and interpretation in this paper, the founding core team and those whose long-term significant effort produced the final CFHTLenS data product.  The second group covers members of the CFHTLenS team who made a significant contribution to the project and/or this paper. CH and LVW co-led the CFHTLenS collaboration.}

\bibliographystyle{mn2e}
\bibliography{lensing}

\appendix
\section{Galaxy models}\label{appendixA}
Galaxy models were created for CFHTLenS according to two basic types, bulge- or disk-dominated, as described
in Section\,\ref{galaxymodels}.  Thus two surface brightness profiles were assumed, for each of the bulge and disk
components of galaxies.  For computational speed, these were two-dimensional distributions with 
radial surface brightness dependencies having S\'{e}rsic index equal to either 1 or 4, which were
given elliptical isophotes by a 2D shear transformation.  The transformation was defined so that the
semi-major axis of a model component was invariant (unlike, for example, the GREAT08 or GREAT10 simulations,
\citealt{bridle10a}, \citealt{GREAT10}).  The rationale for this choice was that the ellipticity of disk galaxies is 
determined primarily by the inclination angle of the disk to the line of sight, in which case the semi-major
axis is the best estimate of an inclination-independent intrinsic size of the galaxy.  
When defining a prior based on the distribution
of intrinsic size of galaxies, that prior should be applied to the distribution of semi-major axes.  

Exponential and de\,Vaucouleurs surface brightness distributions are sharply peaked towards their centres,
so if the surface brightness distributions are generated on a coarse pixel grid, the central pixel
may become incorrectly dominant.  The problem is worse for the de\,Vaucouleurs profile,
$\rho(r) \propto \exp(-7.67 (r/r_{\rm h})^{1/4} )$: the central pixel is a factor 2143 brighter than pixels at
$r=r_{\rm h}$.  A partial solution to the problem is to generate surface brightness distributions on 
a finer-resolution grid, however for moderate oversampling factors 
(a factor 11 oversampling was adopted for CFHTLenS), 
even that is not good enough.  For
CFHTLenS this problem was overcome by applying an integral constraint, requiring the sum of the model pixel
values to equal the expected integral under the de\,Vaucouleurs profile, adjusting the central pixel value
so that this condition was met. An alternative approach would be to use the `photon-shooting' monte-carlo method 
employed in the GREAT simulations.

Model components were convolved with the local PSF pixel-basis model at the location of each galaxy being fitted.  Because
both the galaxy model and the PSF contain Fourier modes above the Nyquist sampling limit of the
observed image, we should seek to generate models at high resolution, convolve with a similarly high-resolution
PSF, and only after that convolution, downsample to the observed sampling (noting that the downsampling
should depend on the galaxy's precise location).  However, as the PSF had to be estimated from the stars
on each exposure, it was not straightforward to generate reliable models with high-frequency modes
correctly represented.  In principle, such high-frequency information may be recoverable from the data
because stars are observed at a variety of positions with respect to the pixel grid, however the PSF
also varies between stars, and each star is itself a noisy measure of the downsampled PSF, so it was 
decided not to attempt to reconstruct oversampled PSF pixel-basis models.  In effect, the PSF models adopted had their
high-frequency modes aliased into the range of observed frequencies.  In an attempt to make an approximate 
correction for that effect, an approximate PSF model was created, assuming circular isophotes, 
which comprised a Moffat profile fitted to the PSF near the centre of the field of view.  The high-resolution
galaxy models were first convolved by the Moffat model, before being downsampled.  This downsampled galaxy model
was then convolved with the actual PSF pixel-basis model for each galaxy, 
but with a multiplicative Fourier-space correction applied to remove the effect
of the Moffat model.  In the limit of band-limited data, an exact convolution by the true PSF model would
have been achieved by this process.  For the actual data, which may contain higher-frequency modes, some
level of correction for those modes was obtained by this approximate method, but clearly a future improvement
would be to reconstruct a high-resolution PSF pixel-basis model 
from the data and deal with the high-frequency modes more rigorously.

It is worth noting that, because the observed stars have been convolved with the detector's 
pixel response function (usually assumed to be a top-hat function), the PSF model employed already included
the effect of the mean pixel response.  No further averaging across pixels was done.

A further problem with generating surface brightness distributions from sheared 2D distributions is that
disk galaxies change their surface brightness profiles with inclination angle: edge-on galaxies have
a surface brightness distribution along their minor axes which are best described as a sech$^2$ function
\citep{vanderkruit81a}. This is thought to be partly a result of the distribution of stars perpendicular
to the disk and partly a result of obscuration associated with the disk \citep{graham08a}.  In order
to attempt to address this, in one set of tests, the 2D disk models were replaced by a 3D model in which
the model surface brightness had a sech$^2$ distribution perpendicular to the plane. Obscuration was
assumed to be distributed the same as the stars (probably an oversimplification of the true obscuration), 
and the projected 2D distribution was obtained by integrating through the 3D distribution.  The obscuration
was set to match the fits reported by \citet{graham08a}.  However, when the tests for systematics,
described in Section\,\ref{systematics}, were made, there appeared to be no significant improvement in the
results. Inspection of typical PSF-convolved 3D models showed that they were indistinguishable from
2D models (once allowance for the apparent increase in observed scalelength had been made, see \citealt{graham08a})
and it was concluded that the extra computational expense of the 3D models was not justified.  The likely
reason for the lack of distinction between 2D and 3D models is that, in the CFHTLenS data, most galaxies
are only marginally resolved, and the semi-minor axis in particular is much smaller than the size of the
PSF.  This test did, however, provide some reassurance that the results would not be sensitive to this
aspect of the galaxy models.  

The 2D models of disk-dominated galaxies that were used here had the 
ellipticity and orientation of the bulge component 
set equal to that of the disk.  The 3D models that were also tested had realistic ellipticity gradients,
caused by the bulge component being less flattened than the disk.  The lack of sensitivity to the
2D or 3D galaxy structure arises because in the CFHTLenS data most galaxies are only marginally resolved,
and hence the bulge ellipticity has little influence on the overall measurement, which is dominated by the
more extended disk component.  In future surveys with better resolution, 
such as the Euclid\footnote{sci.esa.int/euclid}
mission, ellipticity gradients must be taken account of \citep{bernstein10a} and more complex galaxy models
will likely be needed.

\section{Prior distributions}\label{appendixB}
\subsection{Galaxy scalelength}
The disk and bulge scalelength priors are derived from the fits to 
HST WFPC2 observations of galaxies in
the Groth strip by \citet{simard02a}, shown in Fig.\,\ref{fig:simard} for the disk-dominated
galaxies.  The relation between the log of the median major-axis scalelength, $r_d$ and
$i$-band magnitude appears linear.  A least-squares fit to the median values 
in the range $18.5 < i_{814} < 25.5$ yields
\begin{equation}
\log_e\left[r_d{\rm /arcsec}\right] = -1.145 - 0.269(i_{814}-23).
\end{equation}
In applying this relation to CFHTLenS we assumed $i_{814} = i_{\rm CFHT}$.
Those measurements do not strongly constrain the functional
form of the scalelength prior: we adopted a prior of the form 
\[
p(r) \propto r \exp(-(r/a)^\alpha)
\]
where the value of $a$ was adjusted to match the magnitude-dependent median of the \citeauthor{simard02a} fits.
The value of $\alpha$ was also not strongly constrained by those data, so we adopted a value in the
middle of a range that appears to give a good representation of the distribution, $\alpha = 4/3$.
With this choice of $\alpha$, $a = r_d/0.833$.
One concern about the use of the WFPC2 data might be that large, low-surface brightness galaxies may have
been missed owing to the limited surface brightness sensitivity of the observations.  While we cannot
exclude the possibility that some galaxies may be missed, 
we note that the WFPC2 data extend to significantly fainter galaxy magnitudes than
CFHTLenS (Fig.\,\ref{fig:simard}) which mitigates against the possible loss of low surface brightness galaxies
at the CFHTLenS depth. The use of the median galaxy size also guards against the loss of 
low surface brightness galaxies.
HST observations are currently the only way of reliably
probing the size distribution of galaxies on scales $\sim 0.1''$.

The distribution of bulge scalelengths (or equivalent half-light radii) is more problematic.  In principle,
we need to consider the disk- and bulge-dominated populations separately.  The sizes of bulges assumed
in the models for the disk-dominated galaxies have been discussed in Section\,\ref{galaxymodels}.  
The distribution of
sizes of bulge-dominated galaxies might, in principle, be obtainable from the \citet{simard02a}
analysis, however these galaxies are a minority of the population, and given the degeneracies that
exist between scalelength, ellipticity and surface brightness profile, in noisy data, it is difficult to 
obtain a robust estimate of the size distribution with present data. In the local universe, half-light radii
of the most luminous early-type galaxies are about 50\,percent larger than those of late-type galaxies, but at
the galaxy luminosities appropriate for CFHTLenS, $M\sim -19$, they are about a factor 2 smaller \citep{shen03a}.
The distribution of early-type scalelengths 
is not well-known, but we note that the details of the early-type prior are not too important as they only
account for about 10\,percent of the galaxies.
We therefore assign the same
distributions to the half-light radius of the bulge-dominated population as to the exponential scalelength
of the disk-dominated population, so that the half-light radii of early-type galaxies 
are 0.6 times the half-light radii of late-type galaxies, consistent with \citet{shen03a}.
Future work should seek to refine our knowledge of these distributions, particularly beyond the local universe.

\subsection{Galaxy ellipticity}
Fits to faint galaxies do not strongly constrain the distributions of galaxy ellipticity, so
we make the assumption that faint galaxies have the same distribution of disk ellipticity
as low-redshift galaxies of the same luminosity in SDSS: disk ellipticities are determined largely by
the disk's inclination, thickness and optical depth at the waveband of observation, the latter being
luminosity dependent \citep{unterborn08a}.  Accordingly we fit the ellipticity distribution
of 66762 SDSS disk-dominated galaxies selected from DR7 \citep{abazajian09a}
to have g-band absolute magnitude $M_g \le -19$, to correspond in luminosity
and waveband to
galaxies selected at $i' \le 24.5$ in CFHTLenS, for which the median redshift is estimated as $z \simeq 0.66$
\citep{hildebrandt12a}.
The functional form adopted was
\begin{equation}
p(e) = \frac{A e \left(1-\exp\left[\frac{(e-e_{\rm max})}{a}\right]\right)}{ (1+e) (e^2 + e_0^2)^{1/2}} \hspace*{5mm} e < e_{\rm max} 
\label{eqn:edisk}
\end{equation}
where $e \equiv |{\mathbf e}|$ and 
where the maximum ellipticity cutoff $e_{\rm max}=0.804$, obtained from fitting to the distribution,
arises primarily because of the finite thickness of galaxy disks.  The circularity parameter
$e_0 = 0.0256$, also obtained from fitting, reflects the intrinsic non-circular symmetry of
face-on disk galaxies: even face-on galaxies only rarely appear to have circular isophotes.
The fitted value for the dispersion was $a=0.2539$.  The normalisation parameter $A$ was determined
numerically. 

In principle, the ellipticity distribution should be a function of galaxy luminosity.  However, the 
prior distribution only has any significant effect for the lowest signal-to-noise galaxies: 
high signal-to-noise galaxies have well-defined likelihood values and the posterior probability
distribution is largely unaffected by the prior.  Accordingly we chose a galaxy luminosity appropriate
for the faintest galaxies in CFHTLenS selected at the median redshift, and fixed that prior distribution
to be the same for all galaxies.

A prior for the ellipticity distribution of bulge-dominated galaxies is not readily obtained. In the absence of
better information, we
adopted a fit to the ellipticity distribution of bulge-dominated galaxies in \citet{simard02a} of the form
\begin{equation}
p(e) \propto e \exp\left[-b e - c e^2\right]
\label{eqn:ebulge}
\end{equation}
where the best-fit values were $b=2.368$, $c=6.691$ and where the normalisation was determined
numerically.

\begin{figure}
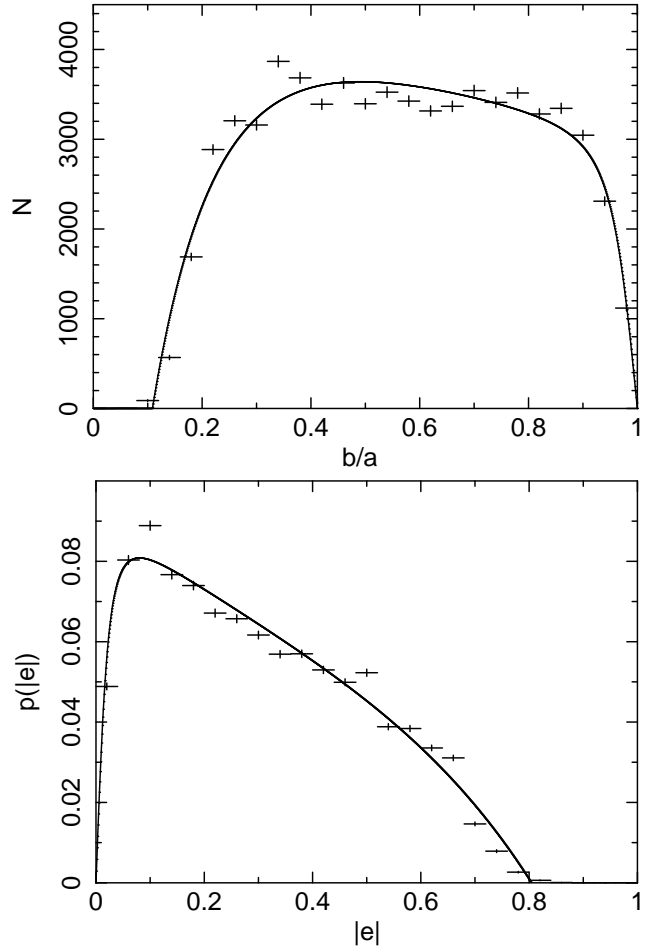

\resizebox{0.48\textwidth}{!}{
\rotatebox{-90}{
\includegraphics{qdist.ps}
}}
\resizebox{0.48\textwidth}{!}{
\rotatebox{-90}{
\includegraphics{edist.ps}
}}
\caption{The observed distributions of axis ratio (upper) and ellipticity (lower) for
disk galaxies with $M_g<-19$ in SDSS, compared with the model function adopted for the
CFHTLenS ellipticity prior (solid curves).
\label{fig:edist}
}
\end{figure}

\subsection{Galaxy bulge fraction}
Likewise, the bulge fraction is not well-constrained by current data, and even in the nearby universe
the bulge fractions deduced for bright galaxies are dependent on the models used to fit them
\citep{graham08a}.  A reasonable distribution to assume, and which is also consistent with
fits to faint galaxy samples such as \citet{schade96a}, is that there are two populations.  Bulge-dominated
galaxies were fit by purely de Vaucouleurs profiles, and account for around 10 percent of the galaxy population. Disk-dominated galaxies were assumed to make up the
remainder, and we assumed the bulge fraction $f = B/T$ to have a truncated normal distribution 
in $f$ with its maximum at $f=0$ and with $\sigma = 0.1$, 
truncated so that $0 \le f < 1$.  The bulge fraction was marginalised over
as described in the main text.
The bulge fraction prior was independent of other parameters.

\subsection{Prior iteration}
In Paper\,II we proposed an iterative scheme to improve the accuracy of the prior probability
distributions using the observed sample of galaxies.  In the STEP simulations the procedure
worked well.  However, when applied to the CFHTLenS data, the iterations did not readily
converge.  It appeared that the main cause of this problem was that in CFHTLenS, unlike
STEP, the galaxies are only marginally resolved, so that there are strong degeneracies between
size, ellipticity and surface brightness profile.  Given that the data are also noisy, it
is not surprising that the iterations tended to converge to arbitrary solutions.  

Fortunately for CFHTLenS, the available HST and SDSS data were sufficient to achieve adequate
estimates of prior distributions.
In future surveys such as Euclid, it may be necessary to use the survey itself to define the
prior distributions.
However, in that survey, galaxies are expected to be rather well resolved, and it is likely
that iterative improvement of the prior would be effective.

\section{Bias in the mean likelihood estimator}\label{appendixC}
Here we discuss the expected bias in the mean ellipticity measured for galaxies 
for a simplified
example of a galaxy with a gaussian surface brightness profile convolved with a 
gaussian PSF.  Because of the unrealistic gaussian surface brightness profiles, 
the results calculated cannot be transferred to the actual CFHTLenS analysis: however, the
gaussian model provides a useful illustration of the origin of bias in the mean ellipticity
estimate.  This work was done independently of that of \citet{refregier12a} and \citet{melchior12a}
but reaches similar conclusions.  The work described here investigates particularly the bias
on the mean likelihood estimator of ellipticity, rather than the maximum likelihood estimator,
and discusses also the correlation with the PSF that is induced by the presence of noise.
Rather than calculate bias in the gaussian model using a Fisher matrix approach, we explicitly
calculate the shape of the likelihood surface for some simple examples.

An elliptical object may be viewed as a sheared version of an isotropic object, and we define the isotropic
coordinates $\mathbf{x'}_i$ of pixel $i$ in terms of the observed coordinates $\mathbf{x}_i$ as 
$\mathbf{x'}_i = \mathbf{A}\mathbf{x}_i$, where 
\begin{equation}
\mathbf{A} = \left(\begin{array}{ll}1-e_1& -e_2\\-e_2 & 1+e_1\end{array}\right)
\end{equation}
and where $e_1$ and $e_2$ are the two components of ellipticity.

We define a gaussian surface brightness profile as
$
f(\mathbf{x}_i) = \exp\left[-\frac{1}{2} \mathbf{x}_i^T C^{-1} \mathbf{x}_i \right]
$
where the pixel surface brightness covariance matrix is given by
\begin{equation}
C = \frac{s^2}{(1-e^2)^2}\left(\begin{array}{cc}1+2e_1+e^2 & 2e_2\\ 2e_2 & 1-2e_1+e^2\end{array}\right)
\end{equation}
where $s$ is a size parameter and
$e = |\mathbf{e}| = \sqrt{e_1^2+e_2^2}$.

If a gaussian galaxy with pixel surface brightness
covariance matrix $C_G$ is convolved with a gaussian PSF with covariance matrix $C_P$ 
then the resulting surface brightness distribution is gaussian with covariance matrix $C_G+C_P$.

If we fit a model surface brightness distribution, $f^m$ to some pixelated data $f$, the log(likelihood) of the fit may be written as
\begin{equation}
\log\mathcal{L} = \frac{1}{2\sigma^2}\frac{\left[\sum_i f^m_i f_i\right]^2}{\sum_i {f^m_i}^2}
\end{equation}
where the sum is over pixels $i$, the background noise has rms $\sigma$ and is assumed stationary, 
the model galaxy's amplitude has been marginalised over
and an arbitrary additive constant has been neglected (Paper\,I).

If the data corresponds to some gaussian galaxy convolved with a gaussian PSF, with independent, stationary 
gaussian noise added to each pixel, we may write
$
f_i = \mathcal{N} \exp\left[-\frac{1}{2}\mathbf{x}_i^T\left(C_G+C_P\right)^{-1}\mathbf{x}_i\right] + n_i
$
for pixel $i$, where $\mathcal{N}$ is a normalising constant and 
$n_i$ is a gaussian random variate representing the pixel noise.  
Neglecting the bias caused by the term $n_i$, the galaxy's profile-weighted signal-to-noise
ratio is given by 
$
\nu_{\rm SN} = \left(\sum_i f_i^2\right)^{1/2}/\sigma
$
which allows us to define the normalisation constant as 
\begin{equation}
\mathcal{N} = \frac{\sigma \nu_{\rm SN}}{\pi^{1/2} |C_G +C_P|^{1/4}}.
\end{equation}

\begin{figure}
\begin{center}
\resizebox{0.48\textwidth}{!}{
\rotatebox{-90}{
\includegraphics{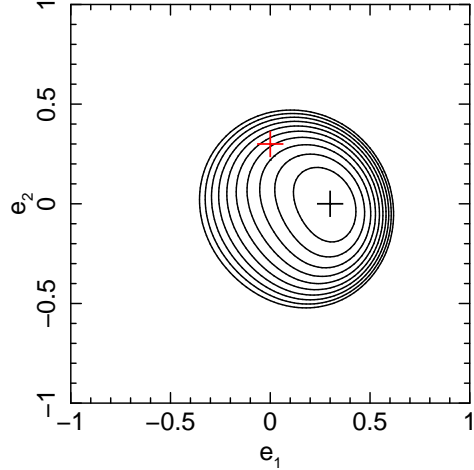}
}}
\end{center}
\caption{Contours of $\log_{\rm e}\mathcal{L}$ for the example gaussian
model and PSF described 
in the text.  The cross at $e_1=0.3, e_2=0$ denotes the ellipticity 
of both the test galaxy
and the maximum likelihood.  The cross at $e_1=0, e_2=0.3$ denotes
the ellipticity of the PSF.  Contours are plotted at intervals of 
0.5 in $\log_{\rm e}\mathcal{L}$ below the maximum,
for a galaxy with signal-to-noise ratio
$\nu_{\rm SN} = 20$.  The contour interval scales as $\nu_{\rm SN}^{2}$.
}
\label{fig:contours}
\end{figure}

Hence the log(likelihood) of the data given some model galaxy with covariance matrix $C_M$, convolved
with the PSF, may be written
\begin{equation}
\log\mathcal{L} = \frac{\left[\sum
\left(\mathcal{N}\mathrm{e}^{-\frac{1}{2}\mathbf{x}_i^T\left(C_G+C_P\right)^{-1}\mathbf{x}_i} + n_i\right)
\mathrm{e}^{-\frac{1}{2}\mathbf{x}_i^T\left(C_M+C_P\right)^{-1}\mathbf{x}_i}
\right]^2}
{2\sigma^2 \sum \mathrm{e}^{-\mathbf{x}_i^T\left(C_M+C_P\right)^{-1}\mathbf{x}_i}}
\end{equation}
and so the expectation value of the log(likelihood) for this particular model $M$ is
\begin{eqnarray}
\left\langle\log\mathcal{L}\right\rangle &=& 
\frac{
2 \pi \mathcal{N}^2 |((C_G+C_P)^{-1}+(C_M+C_P)^{-1})^{-1}|
}
{\sigma^2 |C_M+C_P|^{1/2}} \nonumber \\
& + &
\left\langle\frac{
\left[
\sum n_i f^M_i
\right]^2
}
{\sigma^2 \sum {f^M}_i^2}
\right\rangle 
\end{eqnarray}
where 
$f^M_i = \mathrm{e}^{-\frac{1}{2}\mathbf{x}_i^T\left(C_M+C_P\right)^{-1}\mathbf{x}_i}$
and $\left\langle\sum n_i f^M_i \right\rangle = 0$.
We may use the identity $(A^{-1}+B^{-1})^{-1} = A(A+B)^{-1}B$ in the first term. The second
term has a value of unity, since
$\left\langle\left[\sum f^M_i n_i\right]^2\right\rangle = \sigma^2 \sum {f^M}_i^2$, and ignoring this
irrelevant additive constant, we obtain
\begin{eqnarray}
\left\langle\log\mathcal{L}\right\rangle & = & 
\frac{2\pi\mathcal{N}^2|C_G+C_P| |C_M+C_P|^{1/2}}{\sigma^2|C_G+2C_P+C_M|} \nonumber \\
& = & \frac{2 \nu_{\rm SN}^2 |C_G+C_P|^{1/2} |C_M+C_P|^{1/2}}{ |C_G+2C_P+C_M|} \nonumber\\
& = & 2 \nu_{\rm SN}^2 |(C_G+C_P)^{-1}|^{1/2} |(C_M+C_P)^{-1}|^{1/2} \nonumber \\
& & |(C_G+C_P)^{-1}+(C_M+C_P)^{-1}|.
\label{eqn:mess}
\end{eqnarray}

\begin{figure}
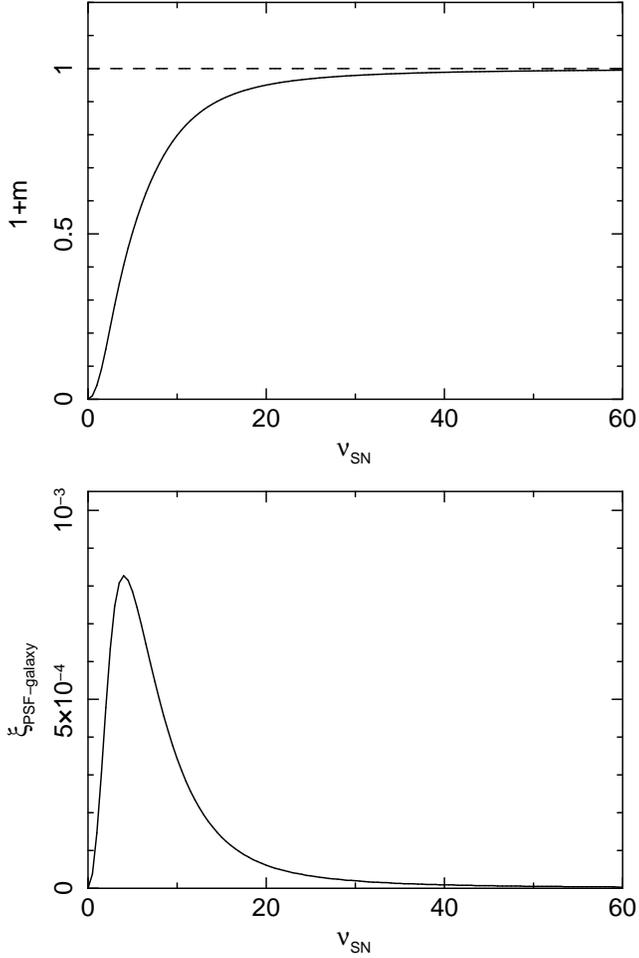

\resizebox{0.48\textwidth}{!}{
\rotatebox{-90}{
\includegraphics{bias.ps}
}}
\resizebox{0.48\textwidth}{!}{
\rotatebox{-90}{
\includegraphics{psfgal.ps}
}}
\caption{
{\em (Upper panel).} The expected shear bias, $1+m$, arising from the bias in the mean ellipticity
for the gaussian model described in the text, as a function of signal-to-noise ratio $\nu_{\rm SN}$.
{\em (Lower panel).} The expected PSF-galaxy cross-correlation signal, $\xi_{\rm PSF-galaxy}$,
in the absence of shear, as a function of $\nu_{\rm SN}$.
}
\label{fig:bias}
\end{figure}

We can repeat this calculation for a set of models $M$, each with a different ellipticity or size, 
and hence build up the 
expected likelihood surface as a function of model ellipticity.  Fig.\,\ref{fig:contours}
shows contours of $\left\langle\log_{\rm e}\mathcal{L}\right\rangle$ for an example where the
galaxy ellipticity was $\mathbf{e}_{\rm g}=\{0.3,0\}$ and the PSF ellipticity was 
$\mathbf{e}_{\rm PSF}=\{0,0.3\}$, with size $\sigma = 3$\,pixels for both galaxy and PSF.
The choice of PSF ellipticity is extreme, much larger than the PSF ellipticity in CFHTLenS,
to better illustrate the bias.  Although the location of the maximum is unbiased, 
the likelihood distribution has a substantial asymmetric distortion, both towards
$e=0$ and towards the PSF ellipticity, which results in the mean ellipticity being
biased.  In these illustrations, the size of the model galaxy has been held fixed (i.e. the galaxy size
is assumed to be known).  If this assumption is relaxed, and the galaxy size is also treated
as a free parameter, the likelihood surface changes shape, with a bias in both the mean and the
maximum likelihood, and an excess of likelihood at extreme ellipticity can be produced
\citep[see also][]{refregier12a, melchior12a}.  We note that the application of a bayesian prior
would suppress that excess likelihood, but we do not consider that effect here.

The upper panel of Fig.\,\ref{fig:bias} shows the shear bias, plotted as $1+m$, expected
in this simple model, for fixed galaxy size, 
for a simplified case where a population of randomly-oriented
gaussian galaxies, all with intrinsic ellipticity $e_{\rm g}=0.25$ and constant shear 
$\mathbf{g} = \{0.05,0\}$,
have been measured with a gaussian PSF, the same size as the galaxy, with 
$\mathbf{e}_{\rm PSF} = \{0,0.1\}$.  The lower panel shows the expected cross-correlation between the
galaxy and PSF.  It can be seen that both measures show significant departures from zero bias
at low signal-to-noise ratio, $\nu_{\rm SN} \la 20$.  In this model, even larger biases are seen for 
gaussian galaxies that are smaller than the PSF.  In CFHTLenS, many galaxies are smaller than the
PSF, however the significant difference in surface brightness profiles of both the galaxies and
the PSF mean that these results cannot be transferred quantitatively to the actual data. Our approach,
described in Section\,\ref{sims}, 
was instead to measure the effects from realistic simulations designed
to match the data.  

\label{lastpage}

\end{document}